\documentclass[12pt]{article}
\usepackage{geometry,float}
\usepackage[parfill]{parskip}
\usepackage{amsmath,amssymb,bbm}
\usepackage{cases}
\usepackage{graphicx,psfrag,epsf}
\usepackage{caption}
\usepackage{subcaption}
\usepackage{tabularx}
\usepackage{microtype}
\usepackage{enumitem}
\usepackage{authblk}
\usepackage{url}
\usepackage[usenames,dvipsnames]{xcolor}
\usepackage[amsmath,amsthm,thmmarks]{ntheorem}
\usepackage{algpseudocode,algorithm}

\usepackage{cleveref}
\crefformat{equation}{(#2#1#3)}
\crefrangeformat{equation}{(#3#1#4) to~(#5#2#6)}
\crefname{equation}{}{}
\Crefname{equation}{}{}
\usepackage[authoryear]{natbib}

\newtheorem{theorem}{Theorem}
\newtheorem{lemma}[theorem]{Lemma}
\newtheorem{example}[theorem]{Example}

\newtheorem{remark}[theorem]{Remark}
\crefname{definition}{\textbf{definition}}{definitions}
\Crefname{definition}{Definition}{Definitions}
\crefname{assumption}{\textbf{assumption}}{assumptions}
\Crefname{assumption}{Assumption}{Assumptions}

\let\hat\widehat
\let\tilde\widetilde

\newcommand{\mP}{\mathbb P}

\newcommand{\mcX}{\mathcal X}
\newcommand{\bH}{\mathbf H}
\newcommand{\bt}{\boldsymbol t}

\renewenvironment{proof}{{\bf Proof.}}{$\Box$}

\usepackage{color}

\newcommand{\blind}{1}

\begin{document}

\def\spacingset#1{\renewcommand{\baselinestretch}%
{#1}\small\normalsize} \spacingset{1}


\if1\blind { \title{\bf Least Ambiguous Set-Valued Classifiers\\ with
    Bounded Error Levels} \author{Mauricio Sadinle$^1$, Jing Lei$^2$,
    and Larry Wasserman$^2$\thanks{Mauricio Sadinle is Assistant Professor in the Department of Biostatistics, 
      University of Washington, Seattle, WA 98195 (e-mail: msadinle@uw.edu).
      Jing Lei is Assistant Professor, Department of Statistics,
      Carnegie Mellon University, Pittsburgh, PA 15213 (e-mail:
      jinglei@andrew.cmu.edu).  Larry Wasserman is Professor,
      Department of Statistics and Machine Learning Department,
      Carnegie Mellon University (e-mail: larry@stat.cmu.edu).
      Mauricio Sadinle was supported by NSF grants SES-11-30706 and
      SES-11-31897.  Jing Lei's research was partially supported by
      NSF grants DMS-1407771 and DMS-1553884.}\\
$^1$\small{Department of Biostatistics, University of Washington, Seattle, WA, USA}\\
$^2$\small{Department of Statistics, Carnegie Mellon University, Pittsburgh, PA, USA}}
		\date{}
  \maketitle
} \fi

\if0\blind
{
  \title{\bf Least Ambiguous Set-Valued Classifiers\\
 with Bounded Error Levels}
  \author{}
	\date{}
  \maketitle
} \fi

\begin{abstract}
In most classification tasks there are observations that are ambiguous
and therefore difficult to correctly label.  Set-valued classifiers
output sets of plausible labels rather than a single label, thereby
giving a more appropriate and informative treatment to the labeling of
ambiguous instances.  We introduce a framework for multiclass
set-valued classification, where the classifiers guarantee
user-defined levels of coverage or confidence (the probability that
the true label is contained in the set) while minimizing the ambiguity
(the expected size of the output).  We first derive oracle classifiers
assuming the true distribution to be known.  We show that the oracle
classifiers are obtained from level sets of the functions that define
the conditional probability of each class.  Then we develop estimators
with good asymptotic and finite sample properties.  The proposed
estimators build on existing single-label
classifiers.  The optimal classifier can sometimes output the empty
set, but we provide two solutions to fix this issue that are suitable for
various practical needs.
\end{abstract}

\noindent%
  \vfill

\newpage
\spacingset{1.45} 

\section{Introduction}
Let $(X,Y)\in \mathcal X\times\mathcal Y$ be a random vector with
unknown distribution $\mP$.  Here $\mathcal X$, typically a subset of
$\mathbb R^d$, is a feature space, and $\mathcal Y=\{1,...,K\}$ is the
label space.  In the traditional multiclass classification problem the
goal is to label each point $x\in\mathcal X$ with the appropriate
class label $y\in\mathcal Y$ in a way that is coherent with $\mP$.
Unlike standard classifiers, which output a single label for each
point in $\mathcal X$, our goal is to construct a set-valued
classifier that assigns a set of {\em plausible} labels to each point
in $\mathcal X$, while assuming that each sample point truly belongs
to a single class.  
Our motivation comes from the fact that in most
classification tasks there are ambiguous observations whose true class
is difficult to determine, yet traditional classifiers are forced to
output single labels.  We argue that assigning sets of plausible
labels provides a better treatment to such instances and therefore
leads to a more informative approach to classification.  While there already exist approaches to set-valued classification, we derive set-valued classifiers that guarantee lower bounds on the probability of containing the true label while minimizing the average number of assigned labels.  

A set-valued classifier is a function $\mathbf{H}:\mathcal X\mapsto
2^{\mathcal Y}$, or in other words, $\mathbf{H}(x)$ is a subset of
$\{1,\ldots,K\}$ for each $x\in \mathcal X$.  We denote the joint
distribution of $(X,Y)$ on $\mathcal X\times \mathcal Y$ by $\mP$.
Having a set-valued output allows us to guarantee levels of confidence
in our predictions.  We consider two types of coverage guarantees:
\begin{center}
\begin{tabular}{ll}
Total           & $\mP\{Y\in \bH(X)\} \geq 1-\alpha,$\\
Class-Specific  & $\mP\{Y\in \bH(X)|Y=y\} \geq 1-\alpha_y\ \ {\rm for\ all\ }y\in {\cal Y},$
\end{tabular}
\end{center}
where we refer to $\alpha$ ($\{\alpha_y\}_{y=1}^K$) as the error level(s), and to $1-\alpha$ ($\{1-\alpha_y\}_{y=1}^K$) as the coverage or confidence level(s).   

Once we fix the desired levels of confidence
there are further properties that we want our classifiers to have.  In particular, we want a classifier that assigns multiple plausible labels to ambiguous observations, but that does so no more than needed.  We therefore would like a classifier $\bH$ with minimal {\em ambiguity},
which we define as
\begin{equation*}
\mathbb{A}(\bH) = \mathbb{E}\{|\bH(X)|\},
\end{equation*}
where $|\cdot|$ is the number of points in a set.  

In Section \ref{sec:oracle} we provide a full characterization of the
optimal set-valued classifiers, which we refer to as LABEL
 (\emph{l}east \emph{a}mbiguous with \emph{b}ounded \emph{e}rror \emph{l}evels).
These optimal classifiers correspond to level sets of the conditional
probability functions $p(y|x)$, that is, they have the form 
$\{y:\ p(y|x) \geq t_y\}$ for some thresholds $\{t_y\}_{y=1}^K$,
where $p(y|x)\equiv \mP(Y=y|X=x)$.  A potentially undesirable property of the optimal classifiers is that they
may lead to empty predictions, that is, $\bH(x)=\emptyset$ for
some points $x\in\mathcal X$, especially when the required coverage
is low.  
We call ${\cal N}=\{x:\ \bH(x) = \emptyset\}$ the {\em null region}.
This region arises because 
minimizing ambiguity can favor making $\bH(x)$ empty, and because some classes may be relatively well separated with respect to the coverage requirements.
We provide solutions to this issue
in Section \ref{ss:null-region}.  

We consider generic plug-in estimators in Section \ref{sec:estimation}, together with a technique called {\em inductive conformal inference} \citep{papadopoulos2002inductive,VovkGS05, ShaferV08, Vovk13} or {\em split-conformal inference} \citep{LeiRW13}, which 
we use to adjust the classifiers to have finite sample,
distribution-free coverage under essentially no conditions.  
We will also see in Section \ref{sec:estimation}, that all of our analyses carry through
even if we let the number of classes $K$ increase with $n$ as long as
$K \equiv K_n = o(\sqrt{n/\log n})$.   In Section \ref{sec:examples} we present data examples that show the advantages of LABEL classifiers.

\subsection{Related work}\label{ss:related}
Classifiers that output possibly more than one label are known as \emph{set-valued classifiers} \citep{Grycko93} or \emph{non-deterministic classifiers} \citep{JoseDB09}.
In another related framework called \emph{classification with reject option} \citep{Chow70,HerbeiW06,BartlettW08,YuanW10,Ramaswamy15}, a classifier may reject to output a definitive class label if the uncertainty is high.
Set-valued classification contains this framework as a special case, as one can
view the ``reject to classify'' option as outputting the entire set of possible labels.
These methods for set-valued classification generally follow the idea of minimizing a modified loss function.  For example, \cite{HerbeiW06}
assigns a constant loss $\rho\in(0,1/2)$ for the output ``reject,'' while
\cite{JoseDB09} defines the loss function as a weighted combination of
precision and recall in an information retrieval framework.
Certain components of such modified loss functions, such as the loss of the output ``reject'' and the weight used to combine precision and recall, lack direct practical meaning and may be hard to choose for practitioners.  

Another line of related work is
\cite{VovkGS05}
and
\cite{ShaferV08},
who introduced a method called
``conformal prediction''
that yields set-valued classifiers
with finite sample confidence guarantees.  In fact, we will show that many of our results are related to those of \cite{Vovketal14,Vovketal16} for conformal prediction.
\cite{LeiRW13,LeiRW11}, \cite{LeiW13}, and \cite{Lei14}
studied the conformal approach
from the point of view of statistical optimality
in the unsupervised, regression and binary classification cases, respectively.
We make use of conformal ideas in Sections \ref{ss:null-region} and \ref{sec:examples}. Recently,
\cite{Hebiri} used asymptotic plug-in methods to derive
classification confidence sets in the binary case.
They control a different quantity, namely, the coverage
conditional on $\mathbf{H}(X)$ having a single element.
Finally, we notice that although it would seem appealing to aim at controlling the conditional coverage $\mP\{Y\in \bH(X)|X=x\} \geq 1-\alpha$, for all $x\in {\cal X}$, which \cite{Vovk13} calls ``object validity,'' Lemma 1 of \cite{LeiW13} unfortunately implies that if $X$ is continuous and $\hat{\mathbf H}$ 
has distribution-free conditional validity,
then $\hat{\mathbf H}$ is trivial, meaning that $\hat{\mathbf H}(x) = \{1,\ldots, K\}$.  

At this point it is important to point out that our contributions do not belong to the literature on ``multi-label'' or ``multi-output'' classification \citep[see, e.g.][]{MLrev,MLKNN}, since in that context each sample point actually has multiple co-occurring characteristics that one wants to jointly predict, such as the presence/absence of multiple diseases in a hospital patient. A  classifier in that context naturally assigns multiple labels to each sample point --- one label associated to each possibly co-occurring characteristic.  In this article, on the contrary, each sample unit truly belongs to only one out of $K$ mutually exclusive classes, but we use a set-valued output to represent the \textit{plausible} classes for each sample unit.

\subsection{Contributions}\label{ss:contributions}

Our framework improves and generalizes the ideas of \cite{Lei14} to the case of $K\ge 2$ classes, where $K$ can even grow with the sample size.  For binary classification, \cite{Lei14} proposed to find two prediction regions $C_y\subset \mcX$, $y=1,2$, as the solution to minimizing $\mP\{X\in C_1\cap C_2\}$ subject to $\mP\{X\in C_y|Y=y\} \geq 1-\alpha_y$, $y=1,2$, and $C_1\cup C_2=\mcX$.  A first difficulty of that approach is that, as stated, the problem cannot be generalized to the multiclass case in a simple manner, and so our extension is technically non-trivial.  
Most importantly, although \cite{Lei14}'s construction seems ideal, the interaction of the problem constraints may lead to multiple solutions, some of which do not provide a meaningful treatment of ambiguous observations.  If we drop the restriction $C_1\cup C_2=\mcX$, the solution to this optimization problem can correspond to regions that do not naturally overlap, thereby leading to a region of empty predictions (null region).  Imposing the constraint $C_1\cup C_2=\mcX$ in such situations leads to multiple solutions, one of which is to fill in the null region with an arbitrary class, which is indeed the solution provided by \cite{Lei14}.  That solution, however, conceals the characteristics of the classification task at hand: the null region arises because the classes are relatively well separated with respect to the coverage requirements.  In other words, in certain classification tasks we may be able to afford higher confidence levels than the ones initially required.  Furthermore, arbitrarily filling in the null region defeats our goal of giving a proper treatment to ambiguous instances, as we illustrate throughout the article, and it is particularly clear in the application to the zip code data in \Cref{ss:ZIP}.  With multiple classes, arbitrarily filling in the null region no longer corresponds to an optimal solution after imposing the constraint $\cup_{y=1}^K C_y=\mcX$ (the excess risk of this approach is characterized in \Cref{thm:excess_risk}).  We therefore provide alternative solutions that give a more appropriate handling of ambiguous instances (Section \ref{ss:null-region}).  Some of our new arguments provide further insights to the problem and lead to significantly more straightforward characterization and estimation of the optimal classifiers. For example, our \Cref{thm:oracle_total,lm:minprob_imp_minamb,th:class_cover}, and their proofs, are very different from the results presented by \cite{Lei14}.

\section{Optimal procedures}\label{sec:oracle}

Our discussion will focus on the case $\mathcal X\subseteq \mathbb R^d$. Let $\mathbb{P}$ denote the joint distribution of $(X,Y)$ on $\mathcal X\times \mathcal Y$.  In this section we derive LABEL classifiers assuming that $\mP$ is known, but in Section \ref{sec:estimation}
we present different estimation procedures.
We assume $\mathbb{P}$ to be absolutely continuous with respect to $\nu(x,y) = \nu_X(x)\nu_Y(y)$ where $\nu_X$ is the Lebesgue measure and $\nu_Y$ is the counting measure.  Let $p$ be the density of $\mathbb{P}$ with respect to $\nu$. 
Throughout the article, we denote $p(x|y)\equiv p_y(x) \equiv p(x|Y=y)$, where
$p(\cdot|Y=y)$ is a density of the conditional distribution of $X$
given $Y=y$, which is assumed to be positive on $\mathcal X$ for each
$y=1,...,K$.  We let $\pi_y \equiv \mathbb{P}(Y=y)$
denote the marginal class probabilities and denote $p(y|x)\equiv \mP(Y=y|X=x)$.  We assume $\pi_y> 0$ for all $y$.  A set-valued classifier $\bH$
can be represented by a collection of sets
\begin{equation*}
C_y = \Bigl\{x\in\mathcal X:\ 
y\in \bH(x)\Bigr\}\,,~~\text{for}~~y=1,...,K.
\end{equation*}
Then, $\bH(x) = \{ y:\ x\in C_y\}$.
Finally, with a little abuse of notation, we can also define $\bH$ as a subset of $\mathcal X\times \mathcal Y$:
$$
\bH = \Bigl\{ (x,y):\ y\in \bH(x)\Bigr\}.
$$
Note that $\bH(x)$ is the $x$-section of $\bH$ and 
$C_y$ is the $y$-section of $\bH$.

\subsection{Total coverage}\label{ss:TotalCoverage}

We start by considering the problem of minimizing the ambiguity
subject to an upper bound $\alpha$ on the total probability of an
error, that is:
\begin{equation}\label{eq::opt1}
\underset{\mathbf{H}}{\min}\ \mathbb E\{|\mathbf{H}(X)|\}\,\quad\text{subject to}\quad
\mP\{Y\notin \mathbf{H}(X)\}\le \alpha.
\end{equation}

\begin{theorem}\label{thm:oracle_total}
Assume that $p(Y|X)$ does not have a point mass at its $\alpha$ quantile, denoted $t_\alpha$. The classifier 
that optimizes \eqref{eq::opt1}
is given by
\begin{equation*}
\mathbf H^*_\alpha = \Bigl\{(x,y): p(y|x)\ge t_\alpha\Bigr\}\,.
\end{equation*}
This optimal classifier can be written as $\mathbf H^*_\alpha(x) = \{y:\ p(y|x)\geq t_\alpha\}$.
\end{theorem}

\Cref{thm:oracle_total} is a consequence of \Cref{lem:NP} by choosing   $f=p(x,y)$ and $g=p(x)$.  
If
$p(Y|X)$ has a point mass at its $\alpha$ quantile, 
 define $t_\alpha=\sup[t: \mP \{p(Y|X)\ge t\}\ge 1-\alpha]$, and $D_\alpha=\{(x,y): p(y|x)=t_\alpha\}$. Then we must have
$\mP(D_\alpha)>0$.
If  $X$ has a continuous distribution then we can choose a subset $D'\subseteq D_\alpha$ and let
$\mathbf H^*_\alpha = \{(x,y): p(y|x)> t_\alpha\}\cup  D'$, with $\mP(\mathbf H^*_\alpha)=1-\alpha$.
If $X$ is discrete, such a subset $D'$ may not always exist, but we can use a randomized rule on $D_\alpha$ as in the original Neyman-Pearson Lemma.
In the rest of this paper we will avoid this complication by assuming
the distribution of $p(Y|X)$ being continuous at $t_{\alpha}$.

\begin{lemma}[Neyman-Pearson]\label{lem:NP}
Let $f$ and $g$ be two nonnegative measurable functions,
then the optimizer of the problem
$$
\min_C \int_{C} g\,\quad\text{subject to}\,\quad \int_C f\ge 1-\alpha,
$$
is given by
$C=\{f/g \ge t\}$
if there exists $t$ such that $\int_{f \ge t g} f = 1-\alpha$.
\end{lemma}

The problem given in expression \eqref{eq::opt1} is equivalent to the one  studied by \cite{Lei14} for the special case of $K=2$, but we do not impose the restriction $C_1\cup C_2=\mcX$.  This constraint leads to optimal arbitrary solutions that may not be meaningful for handling ambiguous instances, as argued in \Cref{ss:contributions}.  Instead in \Cref{ss:null-region} we will provide more principled solutions when the initial classification regions do not cover the whole feature space.  Moreover, the characterization provided in \Cref{thm:oracle_total} is much simpler than that given in \cite{Lei14}.  Recently, \cite{Vovketal16} proved an equivalent result in the context of conformal prediction, where their Theorem 1 is similar to our \Cref{thm:oracle_total}.

Although it seems reasonable to work with procedures that control the
total probability of an error, in some circumstances this approach may
lead to unsatisfactory classifiers.  In particular, when one of the
classes is much more prevalent than the others, the probability of
properly labeling an element of the smaller classes may be quite low,
and it decreases as the probability of the largest class increases.
We illustrate this behavior with the following example.

\begin{example}\label{example:total_coverage}
Consider $\mathcal X=\mathbb R$, $\mathcal Y=\{1, 2\}$, $\mP(Y=1)=0.95$, and the distributions $(X|Y=y)$ being normal with
means $\mu_1=-1$ and $\mu_2=1$, and standard deviations equal to 1.  
In Figure \ref{f:total_coverage_example} we show the densities of the two classes and the specific coverage of each class as a function of the total coverage.  We can see that the probability of correctly labeling an element of class 2 can be quite low,  whereas we would correctly label elements of class 1 with probability almost equal to 1.  
\begin{figure*}[t]
\begin{center}
\includegraphics[width=.45\linewidth]{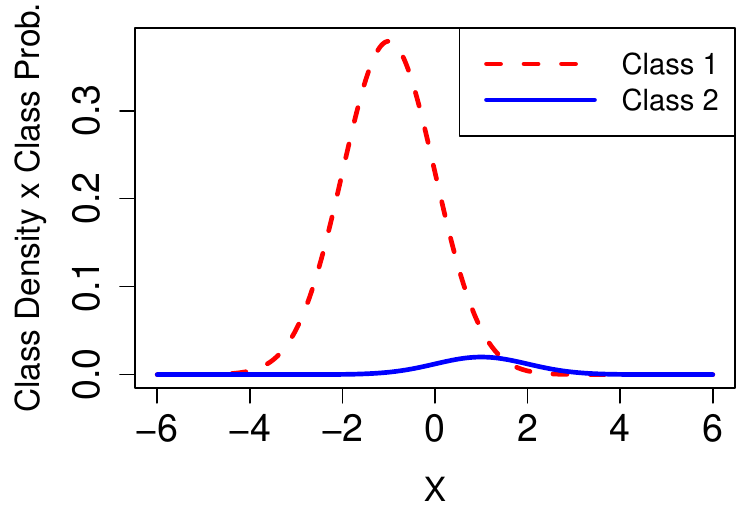}\hspace{1cm}
\includegraphics[width=.45\linewidth]{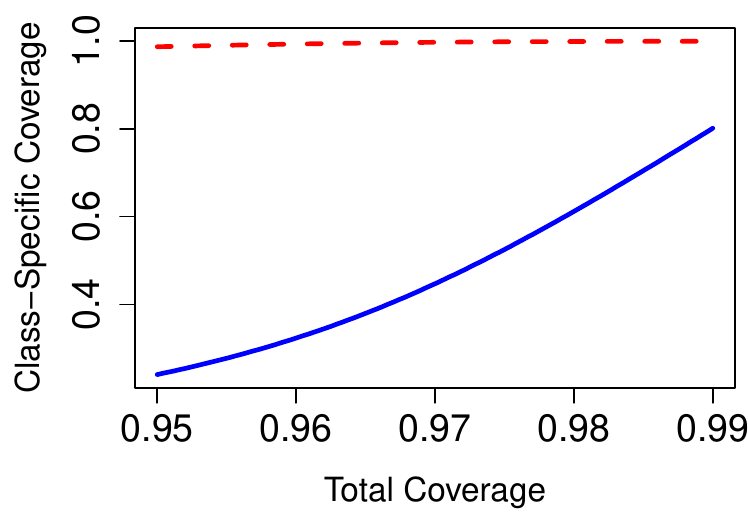}
\end{center}
\begin{minipage}[b]{1\textwidth}
\caption{This figure shows the two classes from Example \ref{example:total_coverage}.  
Left: class-specific densities. Right: class-specific coverage as a function of total coverage.}
\label{f:total_coverage_example}
\end{minipage} 
\end{figure*}
\end{example}
The previous example indicates that a more appropriate
approach should control the coverage of each class, as we show in the
next section.

\subsection{Class-specific coverage}\label{ss:ClassCoverage}

We now derive LABEL classifiers when
controlling the individual coverage of each class.  We consider the following problem:
\begin{equation}\label{eq::opt2}
\underset{\mathbf{H}}{\min}\ \mathbb E\{|\mathbf{H}(X)|\}\,\quad\text{subject to}\quad
\mP\{Y\notin \mathbf{H}(X)|Y=y\} \leq \alpha_y\ \ \ {\rm for\ all\ }y,
\end{equation}
for pre-specified error levels
$\alpha_y$, $y=1,\dots,K$.  Moreover, as we shall see, LABEL classifiers also minimize the probabilities
of incorrect label assignments $\mP\{y\in \bH(X)|Y\neq y\}$ for all $y$.

\begin{remark}\label{rm:amb2}
The ambiguity of a set-valued classifier $\mathbf{H}$ can be expressed as
\begin{align*}
\mathbb{E}\{|\mathbf{H}(X)|\} & = \sum_{y=1}^K \mP\{y\in \mathbf{H}(X)\}.
\end{align*}
\end{remark}

\begin{lemma}\label{lm:minprob_imp_minamb}
If a set-valued classifier $\mathbf{H}$ minimizes the probabilities 
of incorrect label assignments $\mP\{y\in \mathbf{H}(X)|Y\neq y\}$ for all $y\in\{1,\dots,K\}$ 
among all classifiers that have certain error levels $\{\alpha_y\}_{y=1}^{K}$, then it also minimizes the ambiguity.
\end{lemma}

\begin{theorem}\label{th:class_cover}
Given a set of error levels $\{\alpha_y\}_{y=1}^K$,
the set-valued classifier induced by
the sets $C_y=\{x : p(y|x) \ge t_y \}$, with $t_y$ chosen so that $\mP(C_y|Y=y)=1-\alpha_y$,
simultaneously
minimizes the probabilities of incorrect label assignments for all $y$ and
the ambiguity.
\end{theorem}

Theorem \ref{th:class_cover}
tells us 
how to find LABEL classifiers when controlling the error
probability $\alpha_y$ for each class.  A referee pointed out that a similar result had been presented by \cite{Vovketal14} in the context of conformal prediction.
The solution in \Cref{th:class_cover} may
lead to empty predictions, that is, there may exist a region of
$\mathcal X$ where $\bH(x)=\emptyset$.  This phenomenon can also occur
when controlling the total probability of an error as in
Section \ref{ss:TotalCoverage}.  The presence of this null region
occurs when the upper bounds on the error levels are large,
when the classes are well separated, or in practice it could happen if
we have sample points that are anomalies, that is, points that do not
fit any of the existing classes.  In any case, we shall propose principled solutions to cover this region, but we first
illustrate the procedure given by \Cref{th:class_cover} with an
example and defer the discussion on null regions to \Cref{ss:null-region}.

\begin{example}\label{example::3classes}
We consider an example with $\mathcal X=\mathbb R^2$, $\mathcal Y=\{1,
2, 3\}$, $\mP(Y=y)=1/3$ for all $y$, and the distributions $(X|Y=y)$
being bivariate normal with means $\mu_1=(0,3.5)$, $\mu_2=(-2,0)$ and
$\mu_3=(0,2)$, and covariance matrices specified as $\Sigma_1=I_2$,
$\Sigma_2=2I_2$, and $\Sigma_3=\text{diag}(5,1)$, with $I_p$
representing the $p\times p$ identity matrix, and diag representing a
diagonal matrix.
In Figure \ref{f:toy1} we show the classification regions $C_y$
obtained from Theorem \ref{th:class_cover} for different values of
class-specific coverage $1-\alpha_y$, with $\alpha_y=\alpha$ for all $y$, $\alpha=0.2,0.1,0.05$.  We can see that when the required class-specific coverage is not large enough the procedure can lead to a null region.  On the other hand, the null region disappears when the class-specific coverage becomes large and ambiguous classification regions arise as overlaps of the $C_y$ regions.

\begin{figure*}[t]
\centering
\centerline{\includegraphics[width=1\linewidth]{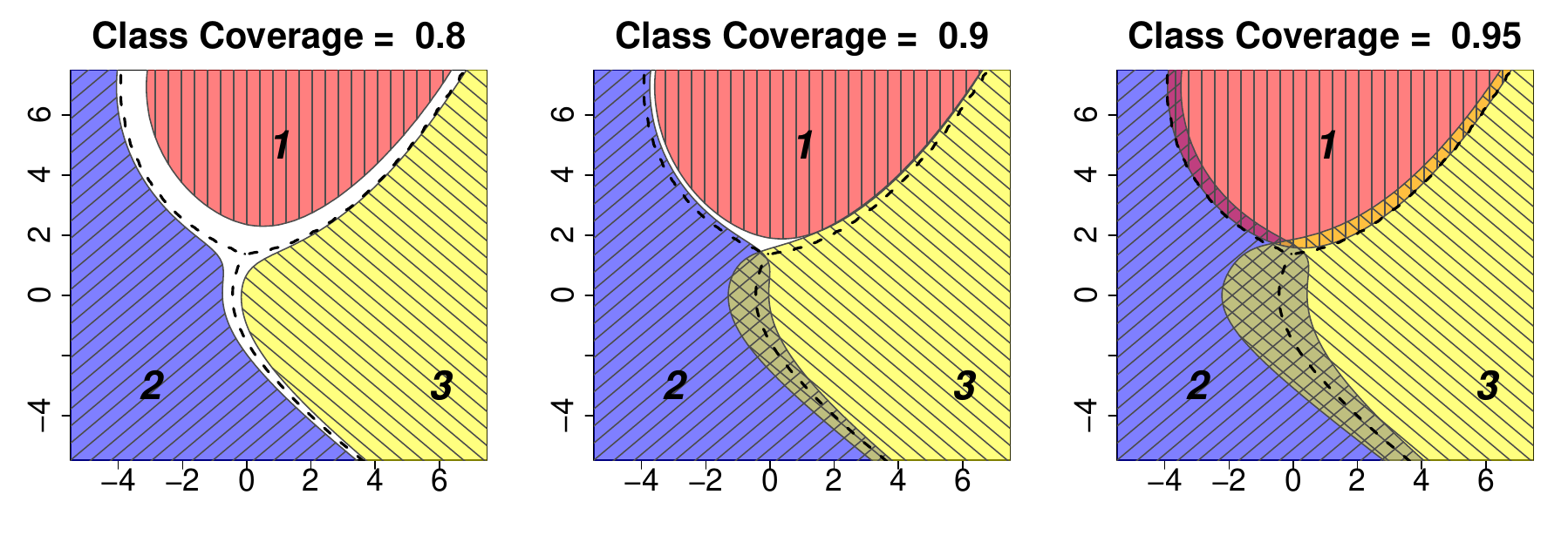}}
\begin{minipage}[b]{1\textwidth}
\caption{This figure shows the three classes from Example \ref{example::3classes}.  For reference, the black dashed lines denote the boundaries of the Bayes classifier's regions. 
Left: Optimal classifier when $\alpha_y=0.2$ for each class.
In this case there is a null region (white) corresponding to $\bH(x) = \emptyset$.
Middle: Optimal classifier when $\alpha_y=0.1$ for each class.
The null region is smaller and an ambiguous region has appeared.
Right: Optimal classifier when $\alpha_y=0.05$ for each class.
The null region is gone but the ambiguity has further increased.}
\label{f:toy1}
\end{minipage} 
\end{figure*}

\end{example}

\section{Dealing with null regions}\label{ss:null-region}

Given a set-valued classification rule $\mathbf H$, the null region is
${\cal N} = {\cal N}(\mathbf H) = \{x:\ \mathbf{H}(x) = \emptyset\}$.  We present two methods
for eliminating these regions of empty predictions.

The first approach, called ``\emph{filling with baseline classifier},'' simply uses a fixed baseline classifier
to fill in the null region.  This method is fast and simple, but it may not properly capture some ambiguous areas of the feature space.  Nevertheless, it provides a simple solution if controlling the nominal error levels is the only concern.

The second method, called ``\emph{accretive completion},'' gradually grows the optimal
classifier by decreasing the class thresholds $t_y$ that define the classes $C_y$ (see Theorem \ref{th:class_cover}),
while minimizing the increments in ambiguity,
until the null region is eliminated.
This approach is more principled and more aligned with the motivation of our framework as it can possibly identify further ambiguous areas inside the null region. We demonstrate this property with the examples presented in \Cref{sec:examples}. 
We recommend this method when thorough detection of ambiguous regions is of concern.

\subsection{Approach I: Filling with a baseline classifier}\label{sss:add_fixed_classifier} 

A simple solution to the problem of the null region is to complete the
set-valued classifier with a given baseline classifier, such as the Bayes
classifier.  Let $h(\cdot)$ be a simple classifier such that $|h(x)|=1$ for all $x$, and define
$$
\mathbf{H}^\dagger(x) =
\begin{cases}
\mathbf{H}(x) & {\rm if\ }\mathbf{H}(x) \neq \emptyset,\\
h(x) & {\rm if\ }\mathbf{H}(x) = \emptyset.
\end{cases}
$$
To justify this method, we start from the following optimization problem that explicitly avoids null region: 
\begin{equation}\label{eq:ambi_alt}
  \underset{\mathbf{H}}{\min}~\mathbb E\{|\mathbf H(X)|\}\,,~~{\rm subject ~ to}~~ \mathbf H(x)\neq \emptyset\,,~\forall~x\,,
  ~~\mP(C_y|y)\ge 1-\alpha_y\,,~\forall~y\,.
\end{equation}
Problem \eqref{eq:ambi_alt} is hard to solve, in general, but the following theorem says that
$\mathbf H^\dagger$ is close to optimal when the null region is small. 

\begin{theorem}[Excess risk bound of $\mathbf H^\dagger$]\label{thm:excess_risk}
Let $\tilde{\mathbb A}$ be the optimal value of problem \eqref{eq:ambi_alt},
 $\mathbf H^*$ a solution to \eqref{eq::opt2}, and $\mathbf H^\dagger$ a classifier such that $|\mathbf H^\dagger(x)|=1$ if $x\in {\cal N}(\mathbf H^*)$ and $\mathbf H^\dagger(x)=\mathbf H^*(x)$ if $x\notin {\cal N}(\mathbf H^*)$ then
  $
  \tilde{\mathbb A}\le \mathbb A(\mathbf H^\dagger) \le \tilde{\mathbb A}+\mP\{\mathcal N(\mathbf H^*)\}.
  $
\end{theorem}

The case of binary classification deserves special attention, given that in that case $\mathbf{H}^\dagger$ always achieves the optimal value $\tilde{\mathbb A}$ of problem \eqref{eq:ambi_alt}.  If the initial LABEL classifier $\mathbf{H}$ leads to a null region, we argue that the regions $C_1$ and $C_2$ have to be  disjoint.  For $C_1$ and $C_2$ to have an overlap there would have to exist $x\in \mathcal X$ such that $p(1|x)\ge t_1$ and $p(2|x)\ge t_2$, which would imply $1=p(1|x)+p(2|x)\ge t_1+t_2$.  But on the other hand, for all elements $x$ in the null region $p(1|x)<t_1$ and $p(2|x)<t_2$, which implies  $1=p(1|x)+p(2|x)<t_1+t_2$.  We conclude that the existence of a null region in the binary case implies $C_1\cap C_2=\emptyset$. In that case, the optimal value $\tilde{\mathbb A}=1$, which is achieved by $\mathbf{H}^\dagger$.  Finally, if the initial LABEL classifier $\mathbf{H}$ does not lead to a null region then $\mathbf{H}^\dagger=\mathbf{H}$, which achieves $\tilde{\mathbb A}$.

It is important to point out that problem \eqref{eq:ambi_alt} can have multiple solutions, some of which may
not necessarily be meaningful when our goal is to properly detect and deal with ambiguous regions, as we explain
in \Cref{example:completing_classifier}.  It is also worth noticing that the excess risk bound of \Cref{thm:excess_risk} also characterizes the procedure that fills in the null region with an arbitrary class.  This motivates the usage of a different approach, as presented in the next subsection.

\begin{example}\label{example:completing_classifier}
Consider the scenario given by \Cref{example::3classes} with 0.8
class-specific coverage.  In this case the regions $C_{y}'$ that are
the optimal solution of problem \eqref{eq::opt2} are all disjoint and
do not cover the whole feature space (left panel of
Figure \ref{f:toy1}). The null region includes cases
that are truly ambiguous, that is, cases where either
$\mP(Y=y|x)\approx 1/3$ for $y=1,2,3$, or $\mP(Y=y|x)\approx 1/2$ for
two values of $Y$, and therefore assigning a single label to such cases 
would not allow us to properly handle their ambiguity.  
Now, notice that by adding the constraint $\bigcup_{y=1}^K
C_y=\mathcal X$ to problem \eqref{eq::opt2} we have that the minimum
value of the ambiguity is 1.  Given that the regions $C_{y}'$ already
achieve the desired levels of coverage, any partition $\{C_y^*\}_{y=1}^K$ of
$\mathcal X$ such that $C_{y}' \subseteq C_{y}^*$ represents an
optimal classifier.  Interestingly, this includes the option of
filling in the null region with an arbitrary class, that is, define
$C_y^*=C_{y}'$ for all $y\neq y_0$ and
$C_{y_0}^*=C_{y_0}'\bigcup \left(\bigcup_{y=1}^K C_{y}'\right)^c$ for some
arbitrary $y_0\in\{1,\dots,K\}$.  We conclude that the problem given in 
Expression \eqref{eq:ambi_alt} may lead to solutions that are not appropriate 
for handling ambiguity in classification.
\end{example}

\begin{remark}
  One can also reduce the chance of  having a null region
  by assigning the same ambiguity loss to the empty set and sets with cardinality one.
  That is, for $\mathbf H(x)\subseteq \mathcal Y$, we define the loss function
  $\ell\{\mathbf H(x)\} = \max\left\{|\mathbf H(x)|-1,~0\right\}, $
   and look for classifiers $\mathbf H$ minimizing $\mathbb E\left[\ell\{\mathbf H(X)\}\right]$, subject
  to coverage requirements.
Theorem 3 of \cite{Vovketal16} provides a characterization, under total coverage, of the optimal classifier among all idealized conformal classifiers
  (classifiers obtained by thresholding some function $g(x,y)$). 
  We note that the loss function $\ell$ does not explicitly eliminate null regions, which is possible
  to occur if the optimal value is $0$.
\end{remark}

\subsection{Approach II: Accretive completion}\label{sss:greedy_grow_classifier}

If a set-valued classifier $\bH$ has a non-empty null region $\mathcal N(\bH)=\{x: \bH(x)=\emptyset\}$,
we call $\bH$ \emph{inadmissible}, otherwise we call $\bH$ \emph{admissible}.  Given $\bt=(t_1,...,t_K)$, denote $\bH_{\bt}=\{(x,y): p(y|x)\ge t_y\}$.
In Theorem \ref{th:class_cover} we showed that, for any $\{\alpha_y\}_{y=1}^K$, the solution 
to the problem \eqref{eq::opt2} is $\bH_{\bt}$ with $\bt$ chosen such that 
$\mP\{Y\in \bH_{\bt}(X)|Y=y\}= 1-\alpha_y$.  Under this solution the inadmissibility of $\bH_{\bt}$ implies $\sum_y t_y>1$.  
To see this, note that
$\bH_{\bt}(x) = \emptyset$ implies that
$p(y|x) < t_y$ for all $y$.
Hence, $1 = \sum_y p(y|x) < \sum_y t_y.$
Therefore,  a sufficient condition for $\bH_{\bt}$ to be admissible is that $\sum_{y=1}^K t_y\le 1$.

Now suppose that given nominal error levels $\{\alpha_y\}_{y=1}^K$, the solution $\bH_{\bt^{(0)}}$, with $\bt^{(0)}=(t_1^{(0)},...,t_K^{(0)})$, to problem \eqref{eq::opt2}
given by Theorem \ref{th:class_cover} is inadmissible. 
 We propose to search for a set of thresholds $\{t_y\}_{y=1}^K$ that has the lowest ambiguity and guarantees admissibility as well as nominal coverage of $\bH_{\bt}$:
\begin{equation*}\label{p:2}
\underset{\bt}{\min}\ \mathbb{E}\{|\bH_{\bt}(X)|\} \text{\ \ subject to\ \ } t_y\le t_y^{(0)}, \forall y;~ \sum_y t_y\leq 1\,.
\end{equation*}
We can proceed in a greedy way to approximate the solution of this problem. The idea is to decrease the thresholds $t_y$, one at a time, in such a way that at each step we achieve the lowest increment in ambiguity.  The detailed procedure is given in \Cref{alg:accretive}.  Notice that for a threshold vector $\bt$ the ambiguity function can be written as
  $\mathbb{A}(\bt)= \sum_{y=1}^K \mP\{p(y|X)\ge t_{y}\}.$

\begin{algorithm}[tb]
\caption{The Accretive Completion Algorithm} 
\label{alg:accretive}
\begin{algorithmic}
    \Require{$\bt^{(0)}=(t_1^{(0)},\dots,t_K^{(0)})$ from the solution of problem \eqref{eq::opt2}, step size $\epsilon$}
     \State $s\gets 0$
     \While{$\sum_y t_{y}^{(s)}>1$}
     \For{$y=1,...,K$ such that $t_{y}^{(s)}-\epsilon\ t_{y}^{(0)}>0$}
		 \State $A_y \gets \mathbb{A}(t_1^{(s)},\dots,t_{y}^{(s)}-\epsilon\ t_{y}^{(0)},\dots,t_K^{(s)})$
      \EndFor
      \State $y^*\gets \arg\min_{y:~t_{y}^{(s)}-\epsilon\ t_{y}^{(0)}>0} A_y$
      \State $\bt^{(s+1)} = (t_1^{(s)},\dots,t_{y^*}^{(s)}-\epsilon\ t_{y^*}^{(0)},\dots,t_K^{(s)})$
			\State $s\gets s+1$
     \EndWhile\\
    \Return{$\bt^{(s)}$}
\end{algorithmic}
\end{algorithm}

\begin{example}\label{ex:greedy}
To build the intuition for the accretive completion procedure consider the first panel of
Figure \ref{f:fill_null_region} (nominal coverage 0.95 for each
class).  We can see that by increasing the coverage of region 1
many points in the null region would be covered by region 1 alone.  On the other hand if we wanted to increase the coverage of
region 3 we would not cover many points in the null region
but we would increase the ambiguity.  This indicates that we can
approximate the solution to the problem by decreasing the
thresholds of the different regions at different rates.

\begin{figure*}[t]
\begin{center}
\includegraphics[width=0.3\linewidth]{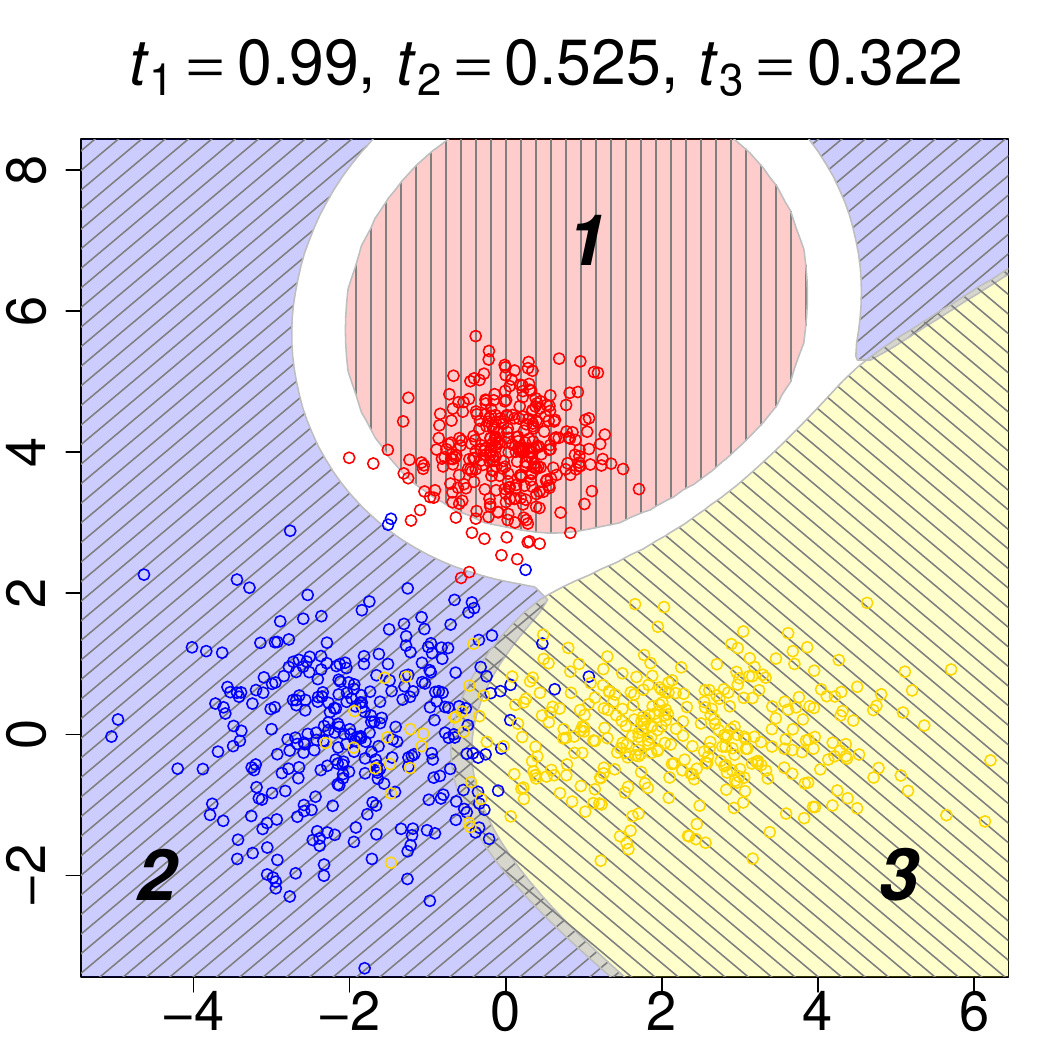}
\hspace{1cm}
\includegraphics[width=0.3\linewidth]{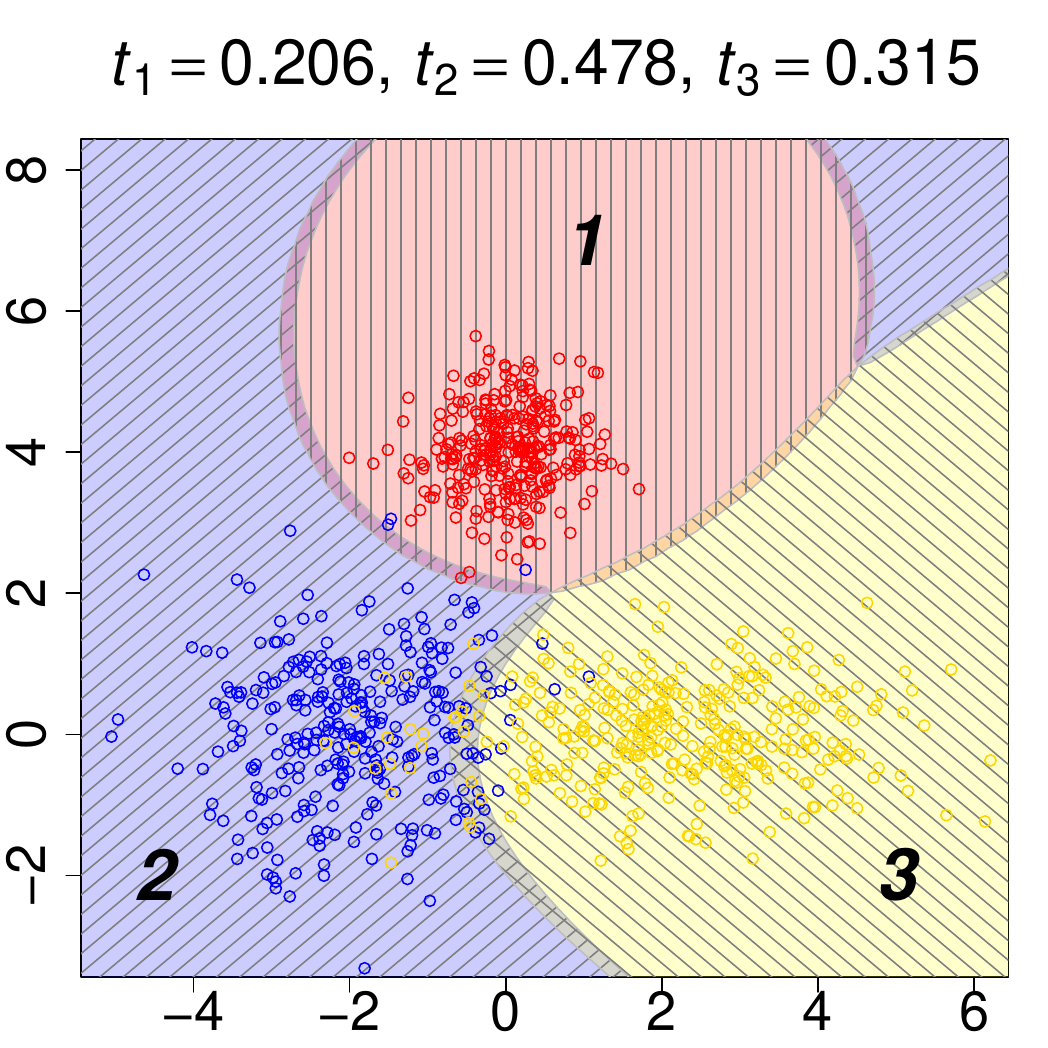}
\end{center}
\begin{minipage}[b]{1\textwidth}
\caption{Left: original solution with class coverage of 0.95. Right: solution of accretive completion algorithm, ambiguity 1.028.  Sample points are drawn to indicate where the probability mass is concentrated. }
\label{f:fill_null_region}\end{minipage} 
\end{figure*}

In Figure \ref{f:fill_null_region} the second panel shows the solution
given by the accretive completion procedure which leads to ambiguity of 1.028.  We can
see that the null region was covered mostly by class 1 since $t_1$
went from 0.99 to 0.206, whereas the other thresholds did not
decrease much.
\end{example}

Although the following remark is obvious from the construction of the accretive completion algorithm, it emphasizes a desirable property of the method.

\begin{remark}
Let $\bH^{+}$ be the classifier obtained from the accretive completion procedure and $\{\alpha^{+}_y\}_{y=1}^K$ its final error levels.  Since the algorithm never increases the thresholds $t_y$, we necessarily have $\alpha^{+}_y\le \alpha_y$, for all $y$, where $\alpha_y$ is the initial error level.  Also, $\bH^{+}$ is the solution to the problem in expression \eqref{eq::opt2} for error levels $\{\alpha^{+}_y\}_{y=1}^K$.  
\end{remark}

Here we note that accretive completion is not necessary in the binary classification case, since the approach of filling with a baseline classifier in the previous subsection already leads to a classifier with the lowest ambiguity, and if we use the Bayes classifier to fill-in the null region we already obtain a classifier with the form $\bH_{\bt}=\{(x,y): p(y|x)\ge t_y\}$, with $t_1=1-t_2$.  

It is also possible to use a similar strategy for growing the total coverage
classifier.
Recall that, in that case,
$\mathbf{H}(x) = \{y:\ p(y|x) \geq t\}$, for a single threshold $t$.
Growing the classifier corresponds to reducing $t$ until there are no null regions.
It is easy to see that this happens when
$t=1/K$. To have the desired coverage, we also need $t\le t_\alpha$, where $t_\alpha$ is the threshold given by Theorem \ref{thm:oracle_total}.
We summarize this as a lemma.

\begin{lemma}
The classifier of the form $\bH_{\bt}=\{(x,y): p(y|x)\ge t\}$ that minimizes the 
ambiguity,
with total coverage at least $1-\alpha$
and empty null-region is given by
$\mathbf{H}(x) = \{y:\ p(y|x)\geq (1/K)\wedge t_\alpha\}$, where
$t_\alpha$ is specified in \Cref{thm:oracle_total}.
\end{lemma}
This lemma suggests that when growing the classifier it
is better to use class-specific coverage as above, especially when $K$ is large.

\section{Estimation and finite sample adjustment}\label{sec:estimation}

Now we consider estimating LABEL classifiers using independent draws $(X_i,Y_i),~i=1,\dots,n$, from $\mP$.
We first study plug-in methods 
which asymptotically mimic the optimal procedures with rates of convergence under standard regularity conditions.
Then we combine these asymptotically optimal procedures with 
\emph{split-conformal inference} to achieve a
distribution-free, finite sample coverage guarantee.

\subsection{The plug-in approach}

The optimal classifiers in the previous sections are level sets of
$p(y|x)$.  There are two things we need to estimate for each
$y=1,\dots,K$: the regression function $p(y|x)$ and the threshold
$t_y$ (when controlling total coverage there is a single threshold
$t_\alpha$).

{\bf The initial estimator of $p(y|x)$.}
Any conventional estimator of $p(y|x)$ can be plugged into the
optimal classifiers.
This could be a direct estimate of $p(y|x)$
or we could estimate
$\pi_y$ and $p_y(x)$ and then use
$p(y|x) = \pi_y p_y(x)/\sum_\ell \pi_\ell p_\ell(x)$.
Here are some specific examples:

\begin{enumerate}
\item[(a)] {\em $k$ Nearest Neighbors ($k$NN)}.
For any $x$, let
$d_k(x)$ be the distance from $x$ to its $k$th nearest neighbor.
Define
$\hat p(y|x) = k^{-1}
\sum_{i=1}^n I(Y_i = y)\ I\bigl\{\|X_i -x\| \leq d_k(x)\bigr\}.$
We ignore cases where several neighbors exist at distance $d_k(x)$, since that occurs with probability zero in our set-up.
We will demonstrate our framework with this estimator in the zip code data example in \Cref{sec:examples}.

\item[(b)] {\em Local polynomial estimator}:  $p(y|x)$ is estimated as
the regression function of $I(Y=y)$ given $X=x$ using the standard local polynomial estimator \citep{Tsybakov09}.
This covers the kernel estimator as a special case:
$
\hat p(y|x) \propto \sum_{i=1}^n I(Y_i = y)\ K_h(x-X_i),
$
where $K_h$ is a kernel with bandwidth $h$.
\item[(c)] {\em Regularized multinomial logistic regression}:
 $ \hat p(y|x) = \exp(\hat\theta_y^T x)/\{1+\exp(\hat\theta_y^T x)\}\,,$
where $\hat\theta_y$ is a possibly penalized logistic regression coefficient
of $I(Y=y)$ on $X$.
\end{enumerate}

{\bf Level sets.}
Having estimated $p(y|x)$ we can determine the cut-off point for the level set according to the
target coverage.

\textit{Case 1: Total coverage.} For the total coverage,
define
$$
\hat{\rm Cov}(t) = \frac{1}{n}\sum_{i=1}^n I\{ \hat p(Y_i|X_i)\ge t\}.
$$
Let
$$
\hat t = \sup\{ t:\ \hat{\rm Cov}(t) \geq 1-\alpha\} = \underset{i}{\max}[~\hat p(Y_i|X_i):~ \hat{\rm Cov}\{\hat p(Y_i|X_i)\} \geq 1-\alpha].
$$
Finally, we take
$\hat{\mathbf{H}}(x) = \{y: \hat p(y|x)\ge \hat t\}$.

\textit{Case 2: Class-specific coverage.} For the class-specific case,
let $\bt=(t_1,\ldots, t_K)$
and define
$\hat C_y = \{ x:\ \hat p(y|x) \ge t_y \}.$
The plug-in estimate of the coverage for class $y$ is 
$$
\hat{\rm Cov}_y(t_y) = 
\frac{\sum_{i=1}^n I( X_i\in\hat C_y) I(Y_i=y)}{\sum_{i=1}^n  I(Y_i=y)}.
$$
Let
$$
\hat t_y = \sup\{ t:\ \hat{\rm Cov}_y(t_y) \geq 1-\alpha_y\} = \underset{i:~Y_i=y}{\max}[~\hat p(Y_i|X_i):~ \hat{\rm Cov}_y\{\hat p(Y_i|X_i)\} \geq 1-\alpha_y].
$$
Finally, we take
$\hat{\mathbf{H}}(x) = \{y: \hat p(y|x)\ge \hat t_y\}$.

{\bf Null region}. If the initial $\hat C_y$'s lead to a null region, we can use a plug-in version of accretive completion (\Cref{alg:accretive}) to cover the whole feature space.  To do this, we replace in \Cref{alg:accretive} each $t_y^{(0)}$ by $\hat t_y$ obtained above, and  $\mathbb{A}(\bt)$  by  
$  \hat{\mathbb{A}}(\bt)= n^{-1} \sum_{i=1}^n \sum_{y=1}^K I\{\hat p(y|X_i)\ge t_{y}\}.$

\subsection{Asymptotic properties}

There are two main assumptions for the development of asymptotic properties of the plug-in level set estimator.
The first one is concerned with the accuracy of $\hat p$.
Assume $\hat p$ satisfies
\begin{equation}\label{eq:initial-est-err}
\mathbb{P}\Biggl\{\sup_x |\hat p_n(y|x)-p(y|x)|\geq \epsilon_n\Biggr\}\leq \delta_n\,,~~\forall~1\le y\le K\,.
\end{equation}
All conventional estimators mentioned in the previous subsection satisfy this sup-norm consistency
under appropriate conditions.

\begin{enumerate}
\item[(a)] $k$NN classifiers: when $p(y|x)$ is Lipschitz on $x$, 
and the distribution of $X$ has intrinsic dimension $d$, then
\eqref{eq:initial-est-err} is satisfied by $k$NN classifiers with
$\delta_n=n^{-1}$ and $\epsilon_n\asymp (\log n/n)^{1/(2+d)}$ when $d\le 2$ ($(\log n/n)^{1/(2d)}$ for $d\ge 3$) \citep{Devroye78}.
\item[(b)] Local polynomial estimators: when $p(y|x)$ is $\beta$-H\"{o}lder smooth, and $\mathcal X$ has dimension $d$,
then \eqref{eq:initial-est-err} holds with $\epsilon_n\asymp (\log n/n)^{\beta/(2\beta+d)}$, $\delta_n\asymp n^{-1}$ for
local polynomial estimator of order $\beta$ \citep{Stone82,AudibertT07,Lei14}.
\item[(c)] Logistic regression estimators: in the case where $d$ is large, if the logistic regression model 
$p(y|x)=\exp(\beta_y^Tx)/\{1+\exp(\beta_y^Tx)\}$ 
holds and $\hat\beta_y$ is estimated with appropriately chosen $\ell_1$ penalty,
then \eqref{eq:initial-est-err} holds with $\epsilon_n\asymp (\log d / n)^{1/4}$ and
$\delta_n\asymp (d^{-1}+\sqrt{\log d /n}\|\beta_y\|_0)$, 
provided that the minimum eigenvalue of $\mathbb E(XX^T)$ is bounded away from $0$ \citep{Geer08}, where $\|\cdot\|_0$ denotes the
number of non-zero entries of a vector. 
\end{enumerate}

The second assumption is on the density of $p(y|X)$ near the cut-off value.  For $1\le y\le K$, let $G_y$ be the distribution function of
$p(y|X)$ with $X$ distributed as $\mP_y$, the conditional distribution of $X$
given $Y=y$. Let $t_y=G_y^{-1}(\alpha_y)$ be the ideal cut-off value
for $p(y|x)$.  Let $G=\sum_{y=1}^K \pi_y G_y$ be the distribution of $p(Y|X)$, with $(X,Y)$ distributed as $\mP$.  

The density condition is, for some constants $\gamma$, $c_1$, $c_2$,
$s_0$,
\begin{equation}\label{eq:margin-condition-class}
c_1|s|^\gamma \le |G_y(t_y+s)-G_y(t_y)|\le c_2|s|^\gamma\,,\ \forall\ s\in [-s_0,s_0]\,,\ 1\le y\le K.
\end{equation}

The corresponding condition for the total coverage is 
\begin{equation}\label{eq:margin-condition-all}
c_1|s|^\gamma \le |G_y(t+s)-G_y(t)|\le c_2 |s|^\gamma\,,\ \forall\ s\in [-s_0,s_0]\,,\ 1\le y\le K.
\end{equation}
The difference is that the threshold $t$ is common for all classes.

\begin{theorem}\label{thm::oracle}
Suppose that \eqref{eq:initial-est-err} and \eqref{eq:margin-condition-class} hold, 
then there exists a constant $c$ such that with probability $1-K\delta_n-n^{-1}$
the plug-in class-specific classifier $\{\hat C_y\}_{y=1}^K$ satisfies
\begin{align*}
\mP_y\left(\hat{C}_{y}\triangle C^*_y\right)\le c\left(\epsilon_n^\gamma+\sqrt{\frac{\log n}{n}}\right)\,,
\end{align*}
where $\mP_y$ is the conditional distribution of $X$ given $Y=y$.
If \eqref{eq:margin-condition-all} holds instead of \eqref{eq:margin-condition-class}, then there exists a constant $c$ such that with
probability $1-K\delta_n-n^{-1}$ the total coverage classifier $\hat\bH$ satisfies
$$
\mP\left(\hat{\mathbf H}\triangle \mathbf H^*\right)\le c\left(\epsilon_n^\gamma+K\sqrt{\frac{\log n}{n}}\right)\,,
$$
where $\mP$ is the joint distribution of $(X,Y)$, and $\mathbf{H}^*$ is the corresponding oracle classifier.
\end{theorem}

\begin{remark}
Suppose we let $K \equiv K_n$ increase with $n$.
Then
$\mP(\hat{\mathbf H}\triangle \mathbf H^*)$
will still go to 0 as long as
$K_n = o(\sqrt{n/\log n})$. 
Thus, our results include the case where the number of classes increases with $n$
as long as it does not increase too fast.
\end{remark}

\subsection{Finite sample coverage via split-conformal inference}\label{ss:split-conf}

Using a method called
{\em split-conformal inference} \citep{LeiRW13} \citep[also known as 
{\em inductive conformal inference} in ][]{papadopoulos2002inductive,Vovk13},
we can guarantee distribution-free, finite sample validity of the classifiers. 

{\bfseries Total coverage.}  The method splits the data in two halves indexed by $\mathcal I_1$ and $\mathcal I_2$.
The first half is used to estimate the conditional probabilities $\hat p(y|x)$, and the second half is used to estimate the distribution of $\hat p(Y|X)$, $(X,Y)\sim\mP$.  Consider augmenting the second half with a hypothetical sample point $(X^*,Y^*)$.  Under the assumption that this new point is drawn independently from $\mP$, the values $\{\hat p(Y^*|X^*)\}\cup\{\hat p(Y_i|X_i)\}_{i\in\mathcal I_1}$ are exchangeable, and so
$$\sigma(X^*,Y^*)\equiv\frac{1}{|\mathcal I_2|+1}\left[\sum_{j\in\mathcal I_2} I \{\hat p(Y_j|X_j)\le\hat p(Y^*|X^*)\}+1\right]$$ 
is uniformly distributed over $\{1/(|\mathcal I_2|+1),2/(|\mathcal I_2|+1),\dots,1\}$.  Therefore, $\sigma(X^*,Y^*)$ can be used to test the hypothesis $H_0:(X^*,Y^*)\sim\mP$.  Given that under $H_0$, $\mP_{\sigma}\{\sigma(X^*,Y^*)\leq \alpha\}\leq \alpha$, for $\alpha\in[0,1]$, we have $\mP_{\sigma}\{\sigma(X^*,Y^*)> \alpha\}\geq 1-\alpha$, and hence any realization of $(X^*,Y^*)$ such that 
$$\sum_{j\in\mathcal I_2} I \{\hat p(Y_j|X_j)\le\hat p(Y^*|X^*)\}>\alpha(|\mathcal I_2|+1)-1,$$ 
would not be rejected as being drawn from $\mP$.  We then need to find 
$$\hat t = \underset{i\in \mathcal I_2}{\min}\left\{~\hat p(Y_i|X_i):~ \sum_{j\in\mathcal I_2} I \{\hat p(Y_j|X_j)\le\hat p(Y_i|X_i)\}>(|\mathcal I_2|+1)\alpha-1\right\},$$
and the set of values in $\mcX\times \mathcal Y$ that would not be rejected is given by
$\{(x,y): ~\hat p(y|x)\ge \hat t~\}\equiv \hat{\mathbf H}$.
It is thus clear that for any distribution, and any $n$,
$\mathbb{P}_{*}\{Y\in \hat{\mathbf{H}}(X)\}\geq 1-\alpha$, where $\mathbb{P}_{*}$ is the distribution of the augmented second half of the sample.

{\bfseries Class-specific coverage.} To guarantee finite sample, distribution-free
validity for class-specific coverage we need to apply the previous method separately to each class.  More specifically, as before, we split the data in two halves indexed by $\mathcal I_1$ and $\mathcal I_2$, and we use the first half to estimate the conditional probabilities $\hat p(y|x)$.  We partition the second half into $K$ groups corresponding to each class, indexed by $\mathcal I_{2,y}=\{i\in\mathcal I_2: Y_i=y\}$, $y\in 1,\dots, K$.  Consider augmenting $\{X_i\}_{i\in\mathcal I_{2,y}}$ with a hypothetical $X^*$, under the hypothesis $H_0:X^*\sim\mP_y$.  If we follow analogous arguments as for total coverage, we obtain that 
if we find 
\begin{equation}\label{eq:hatty_splitconf}
\hat t_y = \underset{i\in \mathcal I_{2,y}}{\min}\left\{~\hat p(y|X_i):~ \sum_{j\in\mathcal I_{2,y}} I \{\hat p(y|X_j)\le\hat p(y|X_i)\}>(|\mathcal I_{2,y}|+1)\alpha_y-1\right\},
\end{equation}
then the classifier $\hat\bH$ obtained from the sets $\hat C_y= \{x: ~\hat p(y|x)\ge \hat t_y~\}$ has class-specific finite sample coverage
$\mathbb{P}_{*}\{Y\in \hat{\mathbf{H}}(X)|Y=y\}\geq 1-\alpha_y\,$, $y=1,\dots,K$, where $\mathbb{P}_{*}(\cdot|Y=y)$ represents the joint distribution of the augmented $\{X_i\}_{i\in\mathcal I_{2,y}}$ sample.  \cite{Vovk13} calls this guarantee ``label validity.''

The accretive completion
(\Cref{alg:accretive}) can be used easily with the split-conformal estimator because one can just apply it to
the second half of the data, while taking $\hat p(y|x)$
as an external input.  To do this, we replace in \Cref{alg:accretive} each $t_y^{(0)}$ by $\hat t_y$ obtained from \Cref{eq:hatty_splitconf}, and $\mathbb{A}(\bt)$
 by
$\hat{\mathbb{A}}(\bt)= |\mathcal I_2|^{-1} \sum_{i\in\mathcal I_2} \sum_{y=1}^K I\{\hat p(y|X_i)\ge t_{y}\}.$

The theoretical results developed in \Cref{thm::oracle} also apply to
the split-conformal estimator, because the key technical ingredients
in the assumption and the proof, such as the sup norm consistency of
$\hat p$ and the empirical distribution of $p(y|X)$, apply to the
split-conformal case.

\begin{theorem}
\label{theorem::inherit}
Let $\{\hat C_y\}_{y=1}^K$ be the split-conformal classifier constructed
from a plug-in classifier $\hat p(y|x)$.   
Then the results of Theorem \ref{thm::oracle}
hold for
$\{\hat C_y\}_{y=1}^K$ under the same conditions, and the classifier has correct distribution-free finite sample coverage.
\end{theorem}

\section{Examples and Comparisons}\label{sec:examples}

In this section we present four examples that illustrate LABEL classification.  
In three of these (Sections \ref{ss:synthetic}, \ref{ss:abalone}, and \ref{ss:ZIP}), the sample sizes are large so we implement the split conformal method.  In Section \ref{ss:iris} the sample size is small so we do not split the data and only study the in-sample performance.  We also demonstrate LABEL classifiers using different base estimators, including kernel regression (Section \ref{ss:synthetic}), $k$NN (Sections \ref{ss:synthetic} and \ref{ss:ZIP}), multi-class logistic regression (Section \ref{ss:iris}), and sparse multi-class logistic regression (Section \ref{ss:abalone}).  We also compare our results with those obtained using the framework of classification with reject option \citep{Chow70,HerbeiW06}, mentioned in Section \ref{ss:related}, where a classifier may reject to output a class label if the correct label is highly uncertain.  The Supplementary Materials contain additional simulation studies comparing the two frameworks.

\subsection{Comparing to classification with reject option}

A classifier with reject option (CWR) is a function $h_R:\mathcal{X}\mapsto \{1,\dots,K,R\}$, with $R$ representing the reject option, which is typically assigned to sample units that are difficult to classify.  The optimal classifier is  derived by minimizing the expected value of the loss function:
\begin{align*}
\ell(y, \tilde y) = \begin{cases}
1, \text{ if } \tilde y\neq y, \tilde y\neq R;\\
\rho, \text{ if } \tilde y=R;\\
0, \text{ if } \tilde y=y.
\end{cases}
\end{align*}
Here $\rho$ should in principle be fixed as the loss of $R$ expressed as a fraction of the loss of a missclassification.  From this loss function  the optimal classifier is
\begin{align*}
h_R^*(x) = \begin{cases}
\underset{y\in \{1,\dots,K\}}{\arg\max} ~ p(y|x), \text{ if } \exists ~ y : p(y|x) \geq  1-\rho;\\
R, \text{ otherwise};\\
\end{cases}
\end{align*}
from which we can define the set-valued classifier
\begin{align}\label{eq:CWR}
\mathbf{H}_R^*(x) = \begin{cases}
\{y\}, \text{ if } h_R^*(x)=y;\\
\{1,\dots,K\}, \text{ if } h_R^*(x)=R.\\
\end{cases}
\end{align}

LABEL classification and classification with reject option are fundamentally different frameworks as they control different quantities.  Nevertheless, we use LABEL classifiers under total coverage for our comparisons.  Given $\hat{p}(y|x)$, a CWR is completely determined by the loss $\rho$ associated to $R$, and therefore to make our comparisons as fair as possible we choose $\rho$ so that $\mathbf{H}_R^*$ in \eqref{eq:CWR} gets the same total coverage of our LABEL classifiers.  Note however that this is not always possible, since with $\rho>1-1/K$ the CWR never outputs $R$.

\subsection{A synthetic nonparametric example}\label{ss:synthetic}

We illustrate LABEL classification using a simulated two dimensional dataset of size $n=2000$ obtained from the distribution presented in Example \ref{ex:greedy}.
We estimate $p(y|x)$ nonparametrically using
$\hat{p}(y|x) = \hat{p}_y(x) \hat \pi_y/\sum_l \hat{p}_l(x) \hat \pi_l$,
where $\hat \pi_y = \sum_i I(Y_i=y)/n$ and
$\hat p_y(x)$ is a  kernel density estimator with bandwidth chosen by Silverman's rule 
\citep{silverman1986density}. 
We use split-conformal prediction as described in \Cref{ss:split-conf}.

\begin{figure}
\begin{center}
\includegraphics[width=1\linewidth]{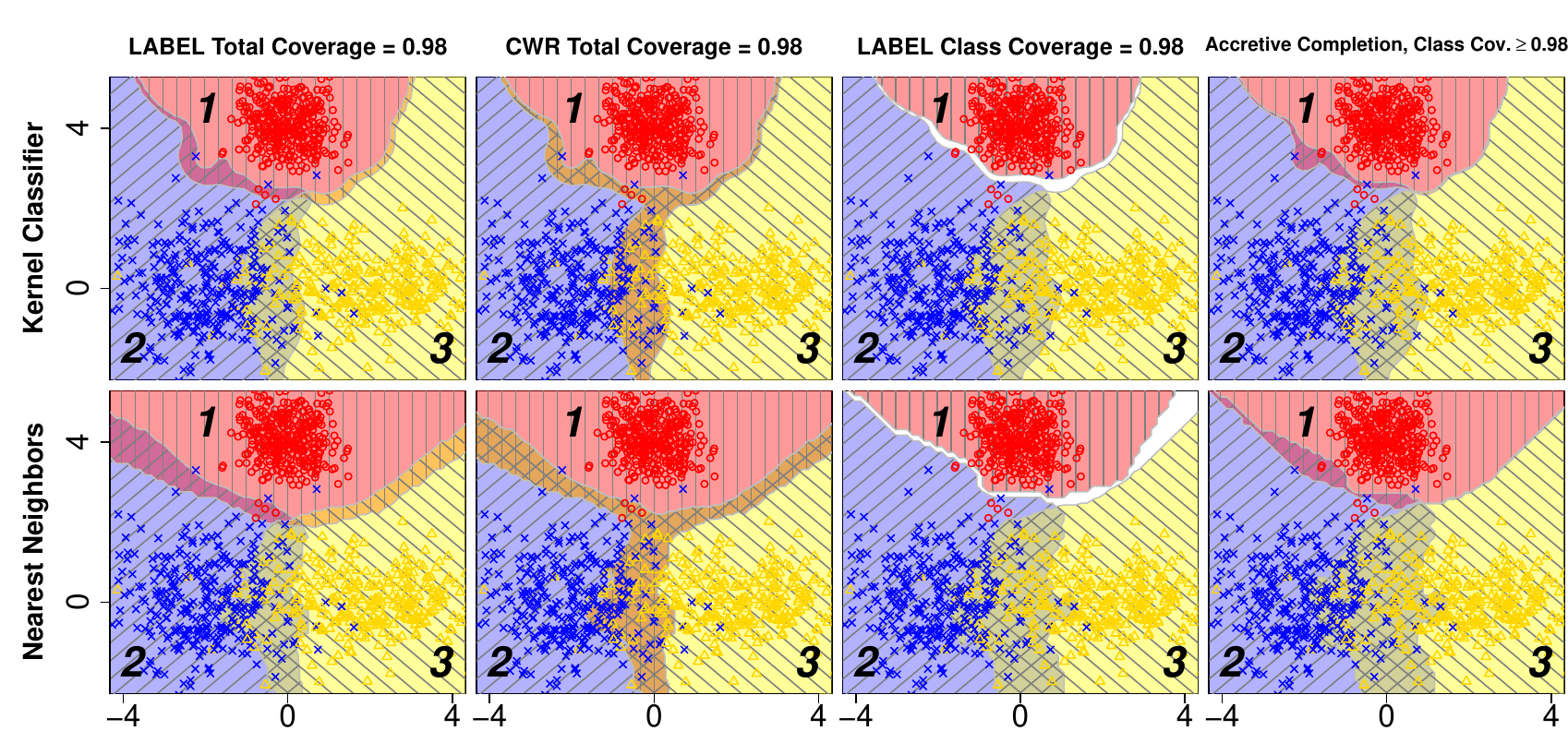} 
\end{center}
\caption{Prediction regions under different set-valued classifiers (columns) using baseline Kernel and nearest neighbors classifiers (rows).  The true data labels are presented as \textcolor{red}{$\circ$} for $Y=1$, \textcolor{blue}{$\times$} for $Y=2$, and \textcolor{Goldenrod}{$\boldsymbol \triangle$} for $Y=3$.
}
\label{fig::kernel}
\end{figure}

The top left plot of Figure \ref{fig::kernel} shows the
LABEL classifier regions $\hat C_y$ ($y=1,2,3$) fixing the total
coverage at 98\%, which leads to an ambiguity of 1.083.  In this case we did not obtain a null region, but this may not be the case with other datasets.  The overlap of the $\hat C_y$ regions represent the areas of the feature space where more than one label is deemed plausible according to this LABEL classifier.  As we had previously argued, controlling the total
coverage can lead to imbalance in terms of the specific coverage of
each class.  In this case the classes 1, 2, and 3 have estimated
specific coverage of 99.7\%, 97.2\% and 97.2\%, respectively.  

In the top row and second column of Figure \ref{fig::kernel} we present the results from CWR choosing $\rho$ so that the set-valued classifier gets 98\% total coverage.  The ambiguity of this classifier is 1.164, and the estimated class coverages are 99.7\%, 97.2\% and 96.9\%.  CWR leads to three regions associated to the single labels 1, 2, and 3, and one general area where the classifier rejects to classify.  From comparing the first two columns of Figure \ref{fig::kernel} we can graphically understand why LABEL classifiers lead to smaller ambiguity:  while a CWR has a general region associated with all possible labels, a LABEL classifier more specifically indicates which subsets of the labels are plausible in different areas of the feature space.  Clearly, the output of LABEL classifiers is more informative as it assigns only the labels that are deemed plausible in each region.

Unlike classification with reject option, LABEL classification also allows us to control the specific coverage of each class.  In
the top row and third column of Figure \ref{fig::kernel} we present the LABEL 
classification regions when fixing the class-specific coverage at
98\%.  Although this approach guarantees the desired coverage for each
class, we obtain a null region despite the large desired coverage due to the fact that class 1 is very well separated.  We therefore use the accretive completion algorithm introduced in
Section \ref{sss:greedy_grow_classifier} to expand the prediction
regions, which are shown in the top right plot of Figure
\ref{fig::kernel}.  The ambiguity of the final classifier is
1.135. 

We also explored the performance of a $k$NN classifier, with $k=15$ chosen by leave-one-out cross validation.  In the bottom row of Figure \ref{fig::kernel} we present plots analogous to those obtained from the LABEL kernel classifier.  We can see that the classification
regions under both approaches are similar in high density areas but
they are dramatically different in low density areas.  The ambiguity
of the final LABEL nearest neighbors classifier is 1.204.

\subsection{Iris data}\label{ss:iris}

\begin{figure}
\begin{center}
\includegraphics[scale=0.35]{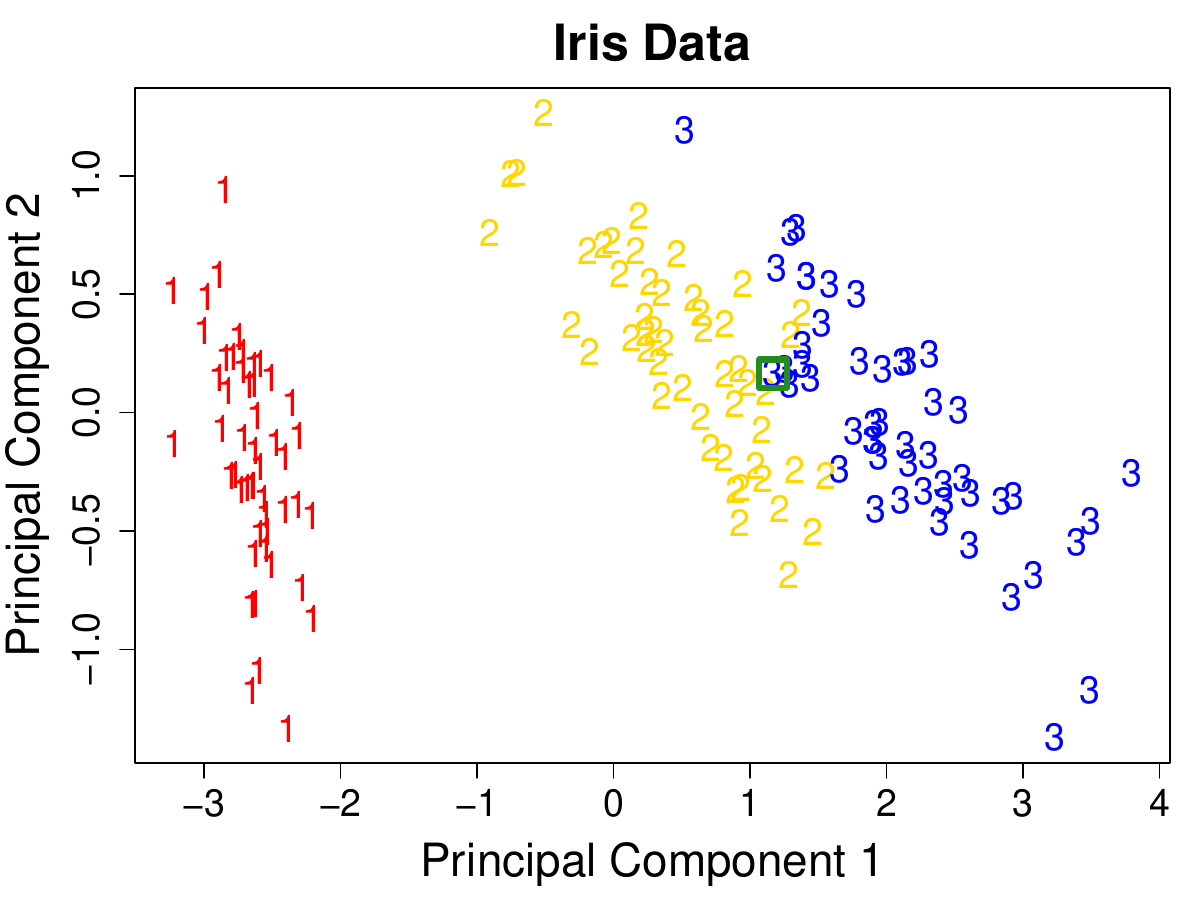} \hspace{.5cm} \includegraphics[scale=0.35]{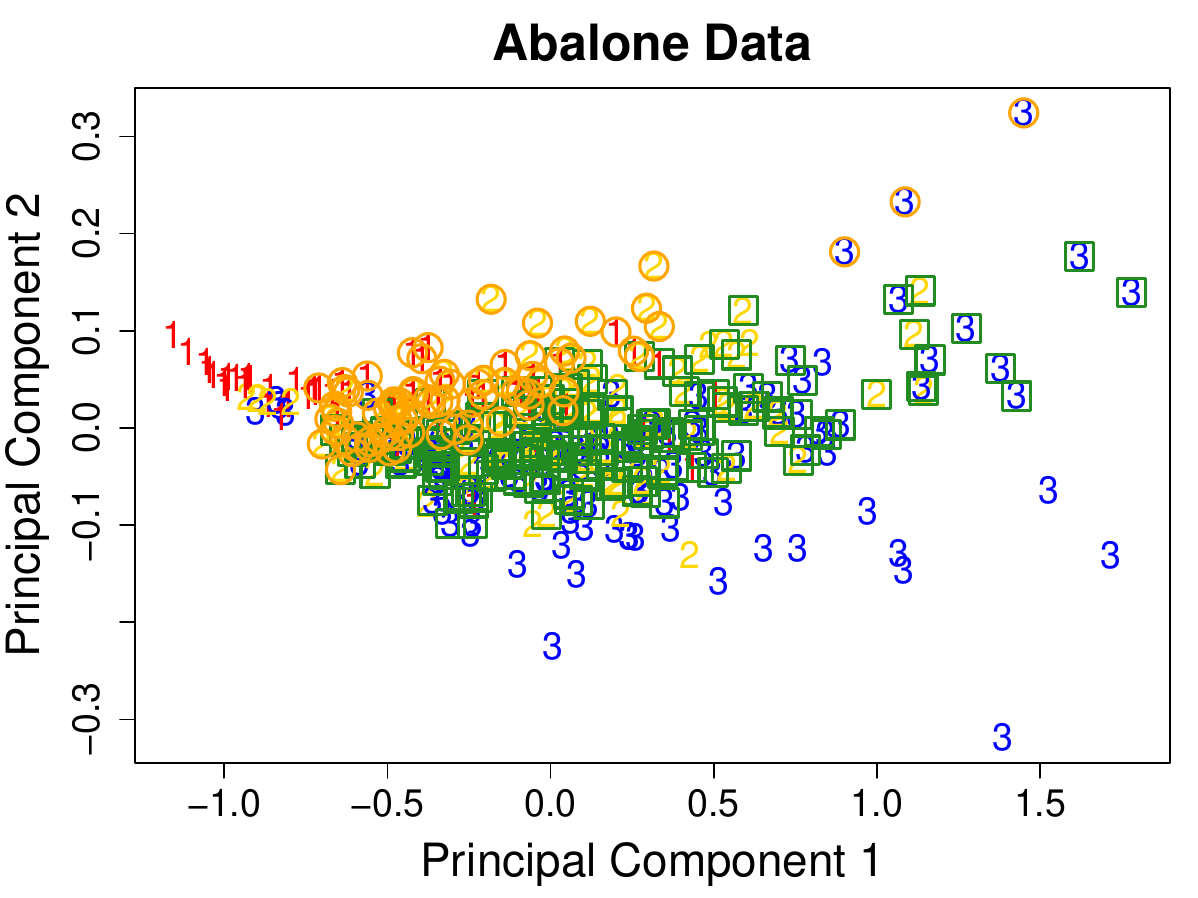}
\end{center}  
\caption{Display of the Iris and Abalone (males) data on their first principal components. Each instance is marked with its true class number.  Ambiguous instances
labeled with $\{1,2\}$ and $\{2,3\}$ by the LABEL classifier are respectively marked with \textcolor{orange}{$\boldsymbol \circ$} and \footnotesize{\textcolor{ForestGreen}{$\boldsymbol \Box$}}.}
\label{fig:iris-pc2}
\end{figure}

The well-known Iris data 
consists of 50 samples from each of three species of Iris
(1: Iris setosa, 2: Iris virginica, and 3: Iris versicolor). 
There are four features:
the length and the width of the sepals and
petals, in centimeters. 
Due to the small sample size we only illustrate the in-sample behavior
of our method, without any sample splitting.
We use a standard multinomial logistic model as the fitting method.
We first construct a LABEL classifier with class-specific coverage
of $98\%$, which leads to class thresholds of 0.999, 0.585, and 0.660, meaning that some regions of the feature space do not have any assigned class label.  We then use accretive completion to eliminate this null region, which decreases the thresholds to 0.376, 0.203, and 0.411, with a final ambiguity of 1.007.  The final classifier assigns single labels to all sample points except for one, which receives labels 2 and 3.  The left panel of \Cref{fig:iris-pc2} shows that this ambiguous instance is indeed
on the boundary between classes 2 and 3.

To analyze these data using a CWR we need to choose $\rho\in [0,2/3]$.  In this case all values in that range lead to a CWR with total coverage greater that 98\%, and taking $\rho>2/3$ leads to a CWR that never outputs $R$.  Therefore with these data no comparable CWR leads to identifying any sample point as ambiguous.

\subsection{Abalone data}\label{ss:abalone}

The abalone data, available from the UCI Machine Learning Repository \citep{Lichman13}, contains measurements of 4177 abalones.
The goal is to predict the age from eight other easy-to-obtain measurements including
sex, length, diameter, height, weight (whole, meat, gut, shell).  To ease the presentation, we grouped the age variable into three categories: 1, young (age 0 to 8); 2, middle (age 9 to 10); 3, old (age 11 and older).  We randomly took one fifth of the data as the testing sample and kept the remainder as the training sample.  For training we used split-conformal inference with a multinomial logistic regression with lasso penalty tuned by ten-fold cross-validation.

The data exhibits substantial overlap between classes, as shown in the right panel of \Cref{fig:iris-pc2}.  An initial LABEL classifier with a target total coverage of 90\% leads to a threshold of 0.217, which means that it does not lead to a null region.  The absence of null region, despite the low initial coverage requirement, is due to the heavy overlap of the classes. The test-sample class-specific coverages are 89\%, 90.7\%, and 92.1\%, with ambiguity of $1.78$.  

A CWR with 90\% total coverage is obtained with $\rho=0.4$, and it leads to test-sample class-specific coverages of 95.5\%, 78\%, and 93.9\%, with ambiguity of $2.02$.  Again, the ambiguity of this classifier is larger than the LABEL classifier, given that a CWR always outputs all labels rather than only the plausible labels for each ambiguous instance.

This example clearly illustrates that controlling total coverage, whether in a LABEL classifier or in a CWR, can lead to great imbalance in the coverage of each class.  We then obtain a LABEL classifier that controls each class-specific coverage at 90\%, obtaining thresholds of 0.21, 0.22, and 0.22, meaning that this classifier does not have a null region, again due to the heavy overlap between the classes.  
In the right panel of \Cref{fig:iris-pc2} we illustrate the results of our final LABEL classifier using the first two principal components of the numeric features of male abalones.  We can see that many instances are ambiguous as the classifier has difficulty distinguishing between young and middle abalones, and between middle and old abalones.  The classifier however does not mix young and old abalones.  The test-sample ambiguity of this final classifier is 1.77.

\subsection{Zip code data}\label{ss:ZIP}

We now analyze the zip code data \citep{LeCunBDHHHJ90},
where the training sample contains 7291 gray scale $16\times 16$ images
of hand-written digits, and the class labels correspond to one of the ten
digits from ``0'' to ``9''.  The class labels are relatively balanced,
with the most frequent digit ``0'' having a proportion of $16.8\%$ and the least
frequent digit ``8'' having $7.6\%$.  The test sample has
2007 images.  The zip code data has been analyzed using a similar framework in \cite{Lei14}, but there the problem is
converted to a binary one in an \textit{ad hoc} manner.  Here we treat
it much more naturally as a multiclass problem and reveal some
interesting features.

\begin{figure}[t!p]
\begin{center}
         \begin{subfigure}[b]{0.32\textwidth}
                 \centering
                 \includegraphics[width=0.98\textwidth]{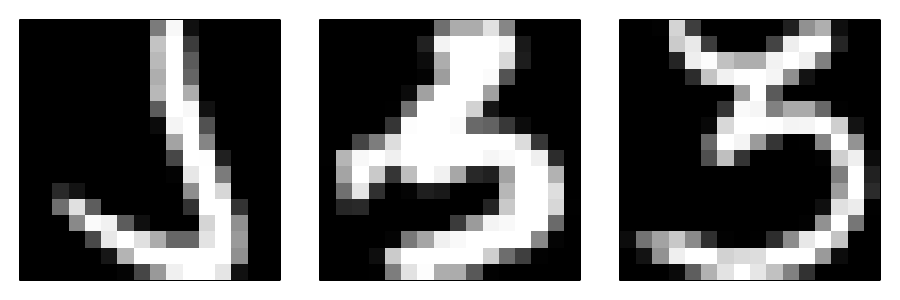}
                 \caption{``3'' and ``5''.}
                 \label{fig:bayes_35}
         \end{subfigure}
     \begin{subfigure}[b]{0.32\textwidth}
             \centering
             \includegraphics[width=0.98\textwidth]{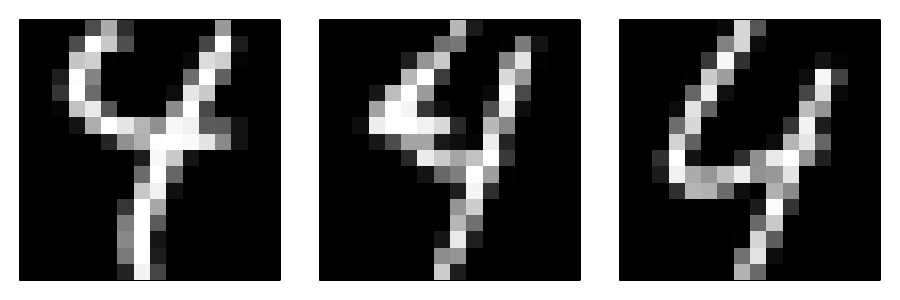}
             \caption{``4'' and ``9''.}
             \label{fig:bayes_49}
     \end{subfigure}
     \begin{subfigure}[b]{0.32\textwidth}
             \centering
             \includegraphics[width=0.98\textwidth]{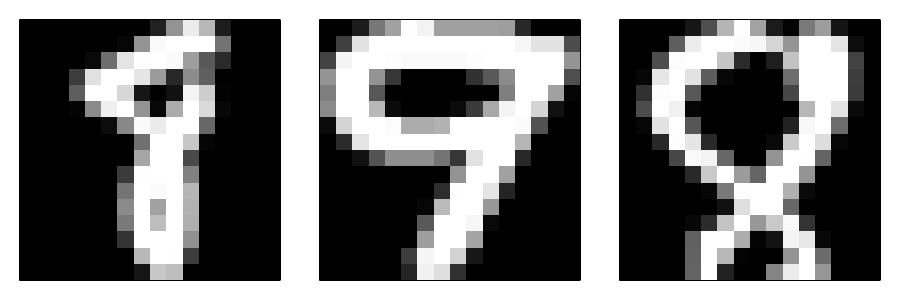}
             \caption{``7'', ``8'' and ``9''.}
             \label{fig:bayes_79}
      \end{subfigure}
\vspace{.5cm}

         \begin{subfigure}[b]{0.32\textwidth}
                 \centering
                 \includegraphics[width=0.98\textwidth]{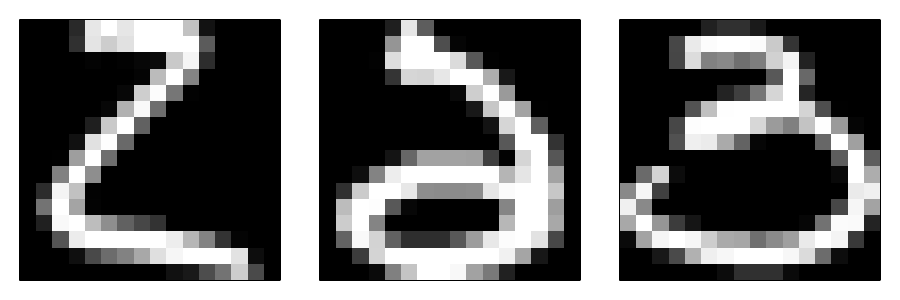}
                 \caption{``0'', ``2'' and ``3''.}
                 \label{fig:add_20}
         \end{subfigure}
     \begin{subfigure}[b]{0.32\textwidth}
             \centering
             \includegraphics[width=0.98\textwidth]{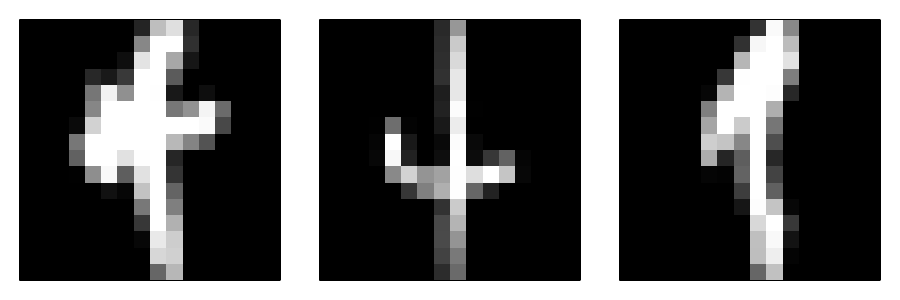}
             \caption{``1'' and ``4''.}
             \label{fig:add_23}
     \end{subfigure}
     \begin{subfigure}[b]{0.32\textwidth}
             \centering
             \includegraphics[width=0.98\textwidth]{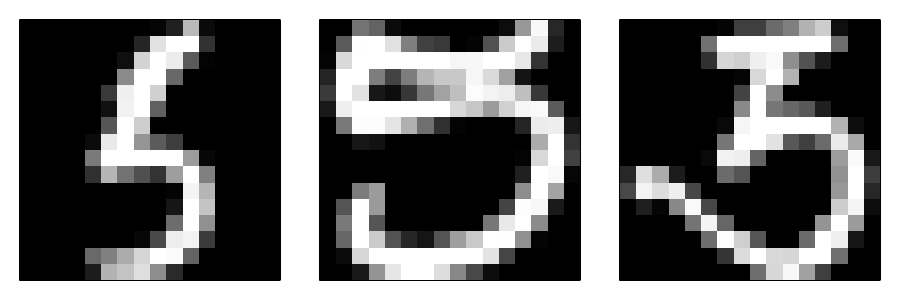}
             \caption{``3'' and ``5''.}
             \label{fig:add_35}
      \end{subfigure}
\end{center}
\caption{Panels (a)--(c): ambiguous images in initial classifier.  Panels (d)--(f): additional ambiguous images obtained after applying the accretive completion algorithm.} \label{fig:ambi}
\end{figure}

We start from a $k$NN classifier with $k=10$ and generate an initial LABEL classifier with target class-specific coverage level of
$0.95$.  We use the split-conformal method described in \Cref{ss:split-conf}, using two thirds of
the training sample to fit the classifier, and the remaining
one third to find the thresholds $\hat t_y$.  The top row of
\Cref{fig:ambi} gives some typical examples of the ambiguous
images in the testing sample using this initial classifier.  Some of the instances in the test sample are such that $\hat{\mathbf
  H}(x)=\emptyset$, given that the thresholds of this initial classifier add up to more than one.  We therefore used accretive completion to remove this null region.  
  The bottom row of
\Cref{fig:ambi} gives examples of images that are marked with multiple labels only after accretive completion.  We can see that many of these images are truly ambiguous, even to the human eye, indicating that filling-in the null region with single label assignments can be potentially misleading.

The test-sample ambiguity of the final LABEL classifier is 1.27.  In \Cref{tab:zip_cooccur} we present a matrix with the co-occurrence of labels in the  test sample according to the final LABEL classifier.  This matrix indicates, for example, that the digits ``3'' and ``5'', ``3'' and ``8'', ``4'' and ``9'', and ``5'' and ``8'' are often hard to tell from each other.  It also tells us that it is easy to tell apart ``0'' from ``7'', ``6'' from ``7'', ``6'' from ``9'', among others, given that no instances receive these label pairs.

\begin{table}
  \begin{center}
    \begin{tabular}{cccccccccccc}\hline
		\multicolumn{10}{c}{Co-occurrence of Label Assignments}\\
		\cline{1-11}
       &\em``0''&\em``1''&``2''&\em ``3''&\em ``4''&\em ``5''&\em ``6''&\em ``7''&\em ``8''&\em ``9''\\
       \hline
      \em``0''&396 & 3 & 36 & 30 & 8 & 35 & 13 & 0 & 13 & 1 \\ 
      \em``1''&3 & 312 & 9 & 4 & 34 & 7 & 4 & 0 & 10 & 2 \\ 
      \em``2''&36 & 9 & 243 & 35 & 20 & 26 & 0 & 10 & 32 & 1 \\
      \em``3''&30 & 4 & 35 & 305 & 7 & \textbf{109} & 0 & 4 & \textbf{70} & 2 \\ 
      \em``4''&8 & 34 & 20 & 7 & 268 & 9 & 4 & 9 & 14 & \textbf{55} \\ 
      \em``5''&35 & 7 & 26 & 109 & 9 & 234 & 9 & 3 & \textbf{43} & 3 \\ 
      \em``6''&13 & 4 & 0 & 0 & 4 & 9 & 171 & 0 & 1 & 0 \\ 
      \em``7''&0 & 0 & 10 & 4 & 9 & 3 & 0 & 192 & 5 & 15 \\ 
      \em``8''&13 & 10 & 32 & 70 & 14 & 43 & 1 & 5 & 214 & 8 \\
      \em``9''& 1 & 2 & 1 & 2 & 55 & 3 & 0 & 15 & 8 & 218 \\ \hline
    \end{tabular}
    \caption{Summary results of applying the LABEL classifier to the zip code data.}\label{tab:zip_cooccur}
  \end{center}
\end{table}

Our final LABEL classifier achieves a test-sample coverage of 0.98, and therefore we build a CWR that achieves the same total coverage.  The CWR that we obtain has a test-sample ambiguity of 2.09.  Similarly as in the previous examples, a CWR leads to a higher ambiguity given that all of its ambiguous outputs have cardinality $10$, whereas our LABEL classifiers are more specific providing sets of plausible labels.

\section{Discussion}
\label{sec::discussion}

LABEL classifiers are the least ambiguous set-valued classifiers that guarantee certain prediction confidence levels.  This framework for classification builds on the strengths of traditional single-valued classifiers, but provides a more informative and principled approach when dealing with ambiguous instances.  As illustrated in Section \ref{sec:examples} and in the simulation studies presented in the Supplementary Materials, we find that LABEL classification provides us with more informative outcomes when compared with classification with reject option, as the former outputs class labels that are plausible for each instance rather than a generic reject option.  LABEL classifiers also give us more control on the desired coverage requirements, as one can control total or class-specific confidence levels.  LABEL classifiers can sometimes output prediction sets that are empty, but we provided different remedies for this problem.

There are many issues that deserve further
investigation.  An important question is how can our consistency results (\Cref{thm::oracle}) be adapted to the modified classifiers obtained from the accretive completion procedure. 
In terms of possible extensions, perhaps the most important one is how to deal with a large number of classes.
In this regard, a promising approach is to organize
the classes into a structure such as a tree.
The tree could be based on prior knowledge
or be discovered from the data.

Finally, we note a possible further enhancement
of our method for class discovery, the idea being to look for new classes in the data.
For example, if we find well defined clusters
in a zone where there is either high ambiguity
or null predictions, then these observations could potentially correspond to a new class.

\section*{Supplementary Materials}

The online supplementary materials contain the proofs of our theoretical results, R code to reproduce the examples in Section \ref{sec:examples}, and additional simulation studies.

\bibliographystyle{apa}
\bibliography{mult_class}

\appendix

\newpage

\section*{Supplementary Materials for ``Least Ambiguous Set-Valued Classifiers with
    Bounded Error Levels''}

\section{Proofs}

We present the proofs of the results that are not clear in the main article. 

{\bfseries \Cref{lm:minprob_imp_minamb}.} \begin{proof}
Let $\mathbf{H}$ and $\mathbf{H}'$ be such that
$
\mP\{Y\in \mathbf{H}(X)|Y=y\} = \mP\{Y\in \mathbf{H}'(X)|Y=y\} = 1-\alpha_y,
$
and 
$\mP\{y\in \mathbf{H}'(X)|Y\neq y\} \geq \mP\{y\in \mathbf{H}(X)|Y\neq y\}$,
for all $y$.  Multiplying this expression by $\mP(Y\neq y)$ we obtain
$
\mP\{y\in \mathbf{H}'(X),Y\neq y\} \geq \mP\{y\in \mathbf{H}(X),Y\neq y\},
$
which can be rewritten as
\begin{align}\label{exp1_lemma1}
\sum_{l\neq y} \mP\{y\in \mathbf{H}'(X)|Y=l\}\pi_l &\geq \sum_{l\neq y} \mP\{y\in \mathbf{H}(X)|Y=l\}\pi_l,
\end{align}
which holds for all $y$.  On the other hand we have
\begin{align*}
\sum_y \mP\{Y\in \mathbf{H}'(X)|Y=y\}\pi_y &= \sum_y \mP\{Y\in \mathbf{H}(X)|Y=y\}\pi_y=
\sum_y (1-\alpha_y)\pi_y.
\end{align*}
Adding Expression \eqref{exp1_lemma1} over all $y$ and combining with the last expression leads to
\begin{align*}
\sum_{y}\sum_{l} \mP\{y\in \mathbf{H}'(X)|Y=l\}\pi_l &\geq \sum_{y}\sum_{l} \mP\{y\in \mathbf{H}(X)|Y=l\}\pi_l,
\end{align*}
which by the law of total probability and Remark \ref{rm:amb2} is equivalent to 
$\mathbb{E}\{|\mathbf{H}'(X)|\}  \geq \mathbb{E}\{|\mathbf{H}(X)|\}$.
\end{proof}

{\bfseries \Cref{th:class_cover}.} \begin{proof}
First, notice that
$\text{logit}\{p(y|x)\}=\log\{p(x|y)/p(x|y^c)\}+\text{logit}(\pi_y)$,
where 
$p(x|y^c)\equiv \sum_{j\neq y}p(x|Y=j)\pi_j/\sum_{j\neq y}\pi_j$. Given that the log and logit functions are monotonically increasing,
this expression implies that the decision regions $C_y$ can
alternatively be based on level sets of the likelihood ratios
$\Lambda_y(x)=p(x|y)/p(x|y^c)$, that is $C_y=\{x : \Lambda_y(x)\ge \ell_y \}$ 
with $\ell_y$ chosen so that $\mP(C_y|Y=y)=1-\alpha_y$.
The region $C_y^c$ therefore corresponds to the Neyman-Pearson
rejection region for testing the null hypothesis $H_0: Y=y$ versus
$H_1: Y\neq y$.  By the Neyman-Pearson lemma we have that the
classifier $\mathbf{H}$ induced by the sets $C_y$ maximizes the
probabilities $\mP\{y\notin \mathbf{H}(X)|Y\neq y\}$, or equivalently
$\mP\{y \in \mathbf{H}(X)|Y\neq y\}$ is minimized.
Finally, by Lemma \ref{lm:minprob_imp_minamb} we 
have that this decision rule $\mathbf{H}$ also minimizes the ambiguity.
\end{proof}

{\bfseries \Cref{thm:excess_risk}.}
\begin{proof}
  Firstly, since $\tilde{\mathbb A}$ is the optimal value of problem 
  \eqref{eq:ambi_alt}, $\tilde{\mathbb A}\le \mathbb A(\mathbf H^\dagger)$.  Now,
 	  $
\mathbb A(\mathbf H^\dagger) = \mathbb E\left[I\{X\in \mathcal N(\mathbf H^*)\}|\mathbf H^\dagger(X)|\right]+
  \mathbb E\left[I\{X\notin \mathcal N(\mathbf H^*)\}|\mathbf H^\dagger(X)|\right]  = \mP\{\mathcal N(\mathbf H^*)\} + \mathbb A(\mathbf H^*), 
  $ 
and the result follows from $\mathbb A(\mathbf H^*)\le \tilde{\mathbb A}$ given that \eqref{eq::opt2} is a relaxation of \eqref{eq:ambi_alt}.
\end{proof}

{\bfseries \Cref{thm::oracle}.}
\begin{proof}
The first part is essentially the same as in \cite{Lei14}.
We prove the second part. 
Let $\hat G_y$ be the empirical distribution of
$p(y|X_{y,1}),...,p(y|X_{y,n_y})$ where $X_{y,1},...,X_{y,n_y}$ are sample points in class $y$.
Let $\hat \mP_y(\cdot)$ be the probability measure corresponding to $\hat G_y$. Define 
$L_y(t)=\{x: p(y|x)\le t\}$, $\hat L_y(t)=\{x: \hat p(y|x)\le t\}$.

We focus on the event
\begin{align*}
E= &  \left\{ \sup_{y,x}|\hat p(y|x)-p(y|x)|\le \epsilon_n\,,\ 
\sup_{y,t} |\hat G_y(t)-G_y(t)|\le c\sqrt{\frac{\log n}{n}}\,,\right.\\
&\qquad
\left.\sup_{y}|\hat \pi_y - \pi_y|\le c\sqrt{\frac{\log n}{n}} \right\},
\end{align*}
which has probability at least $1-K\delta_n-n^{-1}$ if $c$ is chosen large enough and $K$ grows slowly with $n$. Here the first inequality
in $E$ is given by our assumption in \eqref{eq:initial-est-err} and the other two follow from standard empirical process theory.

Recall that for total coverage we use the same threshold for all classes. Let $t^*=G^{-1}(\alpha)$ be the ideal cut-off value for $p(y|x)$.
If $t\le t^*-\epsilon_n-\{(K+1) c c_1^{-1}\sqrt{\log n / n}\}^{1/\gamma}$,
then we have
\begin{align*}
  \hat \mP_y\{\hat L_y(t)\} \le & \hat \mP_y\{L(t+\epsilon_n)\} =\hat G_y(t+\epsilon_n)
  \le G_y(t+\epsilon_n) + c\sqrt{\frac{\log n}{n}}\\
  \le & G_y\left[t^* -\{(K+1) c c_1^{-1}\sqrt{\log n / n}\}^{1/\gamma}\right]+c\sqrt{\frac{\log n}{n}}
  \le G_y(t^*) - cK\sqrt{\frac{\log n}{n}}\,.
\end{align*}
Therefore,
\begin{equation}\label{eq:t-accuracy-1}
  \hat t > t^*-\epsilon_n-\left\{(K+1) c c_1^{-1}\sqrt{\log n / n}\right\}^{1/\gamma}\,,
\end{equation}
because otherwise we have
\begin{align*}
\sum_{y=1}^K \hat \pi_y \hat \mP_y\{\hat L_y(\hat t)\}\le&\sum_{y=1}^K \hat\pi_y \{G_y(t^*) - cK\sqrt{\log n / n}\}\\
\le & \alpha + \sum_{y=1}^K |\hat \pi_y-\pi_y|G_y(t^*) - cK\sqrt{\log n/n} < \alpha\,.
\end{align*}
Similarly we can obtain \begin{equation}\label{eq:t-accuracy-2}
\hat t \le t^*+\epsilon_n+\{(K+1) c c_1^{-1}\sqrt{\log n / n}\}^{1/\gamma}\,,
\end{equation}
and combining \eqref{eq:t-accuracy-1} and \eqref{eq:t-accuracy-2} we have
$
|\hat t - t^*|\le \epsilon_n+\{(K+1) c c_1^{-1}\sqrt{\log n / n}\}^{1/\gamma}\,
$.
(It is worth noting that a rigorous argument of this would require
$\hat p(y|x)$ to have distinct values at the sample points
$X_1,...,X_n$.  This is a minor issue because one can always add very
small random perturbations such as $\hat p(y|X)+\xi$ with $\xi\sim
{\rm Unif}(-n^{-2}, n^{-2})$.)

Then
\begin{align*}
\mP_y\left(\hat{C}_y\backslash C^*_y\right)
=&\mP_y\left\{\hat p(y|X)\ge \hat t,~ p(y|X)< t^*\right\}\\
\le & \mP_y\left[t^*-2\epsilon_n- \left\{(K+1) c c_1^{-1}\sqrt{\log n / n}\right\}^{1/\gamma}\le p(y|X)<t^*\right]\\
\le & c' \left(\epsilon_n^\gamma + K \sqrt{\log n/n}\right),
\end{align*}
for some constant $c'$ depending on $c$, $c_1$, $\gamma$.
Similarly we can obtain
$
\mP_y(C^*_y \backslash \hat{C}_y )\le c'(\epsilon_n^\gamma + K \sqrt{\log n/n})
$,
and
hence
$
\mP_y(\hat C_y \triangle C_y^*)\le c'(\epsilon_n^\gamma+K\sqrt{\log n / n})
$.  Summing over $y$ we have
\begin{equation*}
  \mP\left(\hat{\mathbf{H}}\triangle \mathbf H^*\right)
  =\sum_{y=1}^K \pi_y \mP_y(\hat C_y\triangle C_y^*)
  \le c'\left(\epsilon_n^\gamma+K\sqrt{\log n / n}\right).
\end{equation*}
\end{proof}

\section{Simulation Studies}

\subsection{Univariate Scenarios}

We start with a simple setting to illustrate the fundamental differences between classifiers with reject option (CWRs) and LABEL classifiers.  In this comparison we simulate samples of size $n=4000$, drawing $Y$ from $\{1,2,3\}$ with probabilities that change in three simulation scenarios, summarized in Table \ref{tab:comp_rej_opt}.  We take $X$ to be univariate with distributions $(X|Y=y)$ being normal with means $-2$, 0, and 2, and variances equal to 1.  
We estimate $p(y|x)$ using
$\hat{p}(y|x) = \hat{p}_y(x) \hat \pi_y/\sum_l \hat{p}_l(x) \hat \pi_l$,
where
$\hat \pi_y = \sum_i I(Y_i=y)/n$ and
$\hat p_y(x)$ being a Gaussian density.  We use total coverage of 0.95 for the plug-in LABEL and CWR classifiers.  For each scenario we repeat the simulation 1000 times, and in Table \ref{tab:comp_rej_opt} we report the average ambiguity and class coverages across simulations.

\begin{table}
  \begin{center}
    \begin{tabular}{cccccc}
      \hline
		   & \multicolumn{2}{c}{CWR}& &\multicolumn{2}{c}{LABEL}\\
			\cline{2-3} \cline{5-6}
       Class Probs. &  Ambiguity & Class Coverage & &Ambiguity & Class Coverage\\
      \hline
      $(.45,.10,.45)$ & 1.28 & $(1.00, 0.53, 1.00)$ & & 1.21  & $(0.98, 0.64, 0.98)$\\
			$(.33,.33,.34)$ & 1.94 & $(0.98, 0.89, 0.98)$ & & 1.51  & $(0.96, 0.93, 0.96)$\\
			$(.60,.30,.10)$ & 1.78 & $(0.99, 0.89, 0.92)$ & & 1.41  & $(0.98, 0.91, 0.88)$\\ 
      \hline
    \end{tabular}
    \caption{Comparing LABEL classifiers with total coverage control to CWRs in a three-class problem. Simulation details are given in the main text.  Quantities reported here are averages over 1000 simulation replicates.}\label{tab:comp_rej_opt}
  \end{center}
\end{table}

From Table \ref{tab:comp_rej_opt} we can see that, across the three scenarios that we considered, the LABEL classifier has smaller average ambiguity than the CWR.  In fact, the ambiguity of the LABEL classifier was smaller in all 1000 simulation replicates within each simulation scenario, not only on average.  This is rather natural, since in our construction of LABEL classifiers assigning the output $\{1,2,3\}$ to a sample point is penalized more than an output containing only two labels.  CWRs effectively assign the output $\{1,2,3\}$ to all ambiguous sample points, and therefore do not specify which labels are plausible for a given instance.  In the scenarios explored in our simulation study, classes 1 and 3 are relatively well separated, and therefore the LABEL classifier assigns the outcomes $\{1,2\}$ and $\{2,3\}$ to the sample points in the overlap of classes 1 and 2, and 2 and 3, respectively.  To such cases, the CWR assigns $\{1,2,3\}$, which is less informative and leads to larger ambiguity.  LABEL classifiers are therefore more informative in reporting ambiguous cases as they indicate the set of plausible labels for each instance.

We also conclude from Table \ref{tab:comp_rej_opt} that controlling total coverage can lead to very uneven class coverage with both LABEL classifiers and CWRs.  Notwithstanding, our framework can also be used to control class-specific coverage, something that cannot be done using CWRs.

\subsection{Data-Based Multivariate Scenario}

To explore the performance of LABEL classifiers in comparison with CWRs under more complex scenarios, we use the Abalone data analyzed in Section \ref{ss:abalone} to create a synthetic population from which we simulate.   The goal here is to construct a synthetic population based on real data, not necessarily study the repeated sample performance of the analysis in Section \ref{ss:abalone}.  We start by creating $K=5$ classes using the age variable (unlike in Section \ref{ss:abalone}, where we took $K=3$ to ease the presentation).  We then use the seven numeric features in the Abalone data  to obtain a mean vector and a covariance matrix within each class.  The population is then defined as a five-component mixture of seven-dimensional normal distributions, with means, covariance matrices, and population proportions obtained from the abalone data.  

Each of the 1000 replicates in this simulation study consists of  $n = 4000$ draws from the aforementioned mixture.  The estimator of $p(y|x)$ is a simple multinomial logistic regression.  We controlled the total coverage of the classifiers at 0.95.  For CWRs this means that we chose the cost of the reject option $\rho$ so that its total coverage is greater or equal to 0.95.

\begin{table}[t]
  \begin{center}
    \begin{tabular}{ccc}
      \hline
        &  Ambiguity & Class Coverage \\
      \hline
CWR & 3.64 & $(0.87, 0.98, 0.93, 0.92, 0.97)$ \\
LABEL & 1.87 & $(0.90, 0.99, 0.97, 0.69, 0.53)$ \\\hline
    \end{tabular}
    \caption{Comparing LABEL classifiers and CWRs under total coverage control in a multivariate problem. Simulation details are given in the main text.  Quantities reported here are averages over 1000 simulation replicates.}\label{tab:comp_rej_opt2}
  \end{center}
\end{table}
\begin{figure}[h]
\begin{center}
\includegraphics[scale=0.7]{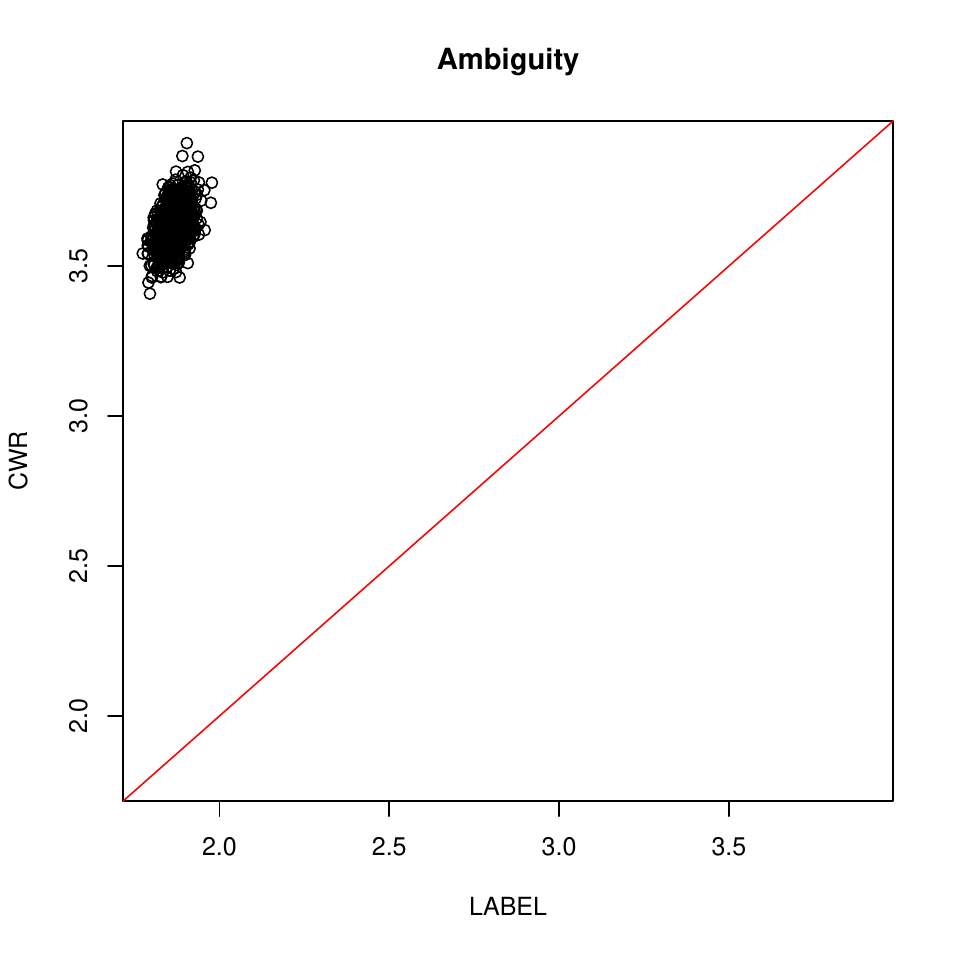}
\end{center}  
\caption{Comparison of ambiguity obtained from LABEL and CWR across 1000 simulation replicates in multivariate scenario.}
\label{fig:sim_ambi_comp}
\end{figure}

In Table \ref{tab:comp_rej_opt2} we compare the ambiguity and class coverages that we obtain on average from LABEL and CWR.  Similarly as with the other examples and simulation scenarios, LABEL classifiers provide smaller ambiguity values, meaning that these are more informative than CWRs as the latter provide larger outputs.  Indeed, in Figure \ref{fig:sim_ambi_comp} we show that LABEL ambiguity is smaller than CWR ambiguity in all simulation replicates.  

In Table \ref{tab:comp_rej_opt2} we can also see that LABEL classifiers tend to give very imbalanced class coverages.  In this simulation scenario this occurs because three of the classes have very small class probabilities (classes 1, 4 and 5), and in addition two of these are not very well separated from a third one (classes  4 and 5 are very close to class 3).  This phenomenon was  illustrated in Example \ref{example:total_coverage}.  Regarding CWRs, their larger outputs lead in this case to higher and more balanced class coverages given that instances receive all labels when they get assigned the reject option.  Nevertheless, balance of class coverage cannot be guaranteed with CWRs, whereas LABEL classifiers can indeed be used under class-specific coverage control.

\end{document}